%% file: final.tex
\documentclass[english]{programming}
\pdfoutput=1

\usepackage[backend=biber]{biblatex}
\usepackage{forloop}

\usepackage{multicol}
\lstdefinelanguage[programming]{TeX}[AlLaTeX]{TeX}{%
  deletetexcs={title,author,bibliography},%
  deletekeywords={tabular},
  morekeywords={abstract},%
  moretexcs={chapter},%
  moretexcs=[2]{title,author,subtitle,keywords,maketitle,titlerunning,authorinfo,affiliation,authorrunning,paperdetails,acks,email},
  moretexcs=[3]{addbibresource,printbibliography,bibliography},%
}%
\newcommand*{\CTAN}[1]{\href{http://ctan.org/tex-archive/#1}{\nolinkurl{CTAN:#1}}}

\begin{document}

\title{What Is the Best Way For Developers to Learn New Software Tools?}
\subtitle{An Empirical Comparison Between a Text and a Video Tutorial}

\author{Verena K\"afer}
\authorinfo[content/pics/verena_kaefer]{studied computer science at the University of Stuttgart where she also started with her doctorate in May 2016. Her research interests include tutorials for computer software and behavioral software engineering. Her ORCID is 0000-0002-7070-4519. Contact her at \email{verena.kaefer@informatik.uni-stuttgart.de}. She designed and conceived the experiments, performed them, analyzed the results and wrote the paper.}
\author{Daniel Kulesz}
\authorinfo[content/pics/kuleszdl_2016]{is a Ph.D. student and researcher at the University of Stuttgart. His main research interests are quality assurance
of spreadsheets and end-user programming. Besides that, he is also an active software engineering practitioner in industry and lecturer. He is a member of ACM SIGSOFT and fellow of the Free Software Foundation Europe. He designed and conceived the experiments, performed them and wrote
the paper. Contact him at \email{daniel.kulesz@informatik.uni-stuttgart.de}.}
\author{Stefan Wagner}
\authorinfo[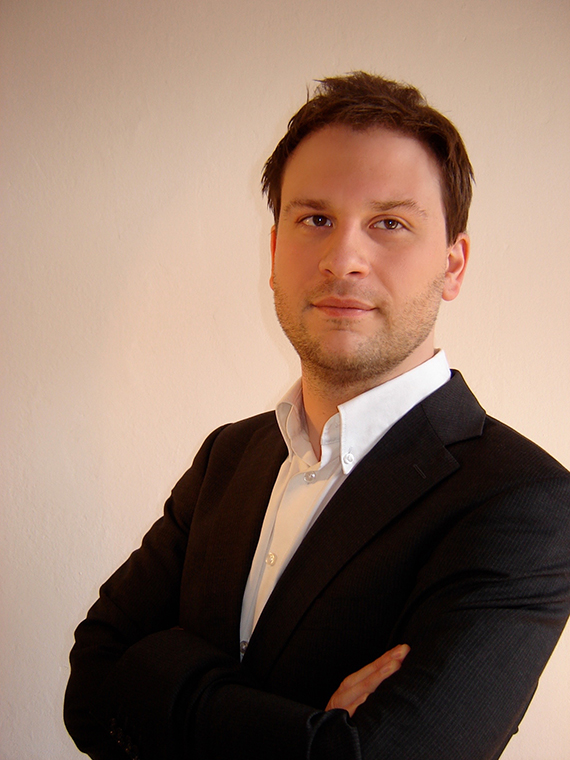]{is full professor of software engineering at the University of Stuttgart. He studied computer science in Augsburg
and Edinburgh and holds a doctoral degree in computer science from the Technical University Munich. His research interests include requirements
engineering, software quality, safety and security engineering, agile software development and empirical/behavioural software engineering. He is
a member of GI, ACM and the IEEE Computer Society. His ORCID is 0000-0002-5256-8429. Contact him at \email{stefan.wagner@informatik.uni-stuttgart.de}.
He contributed to the experiment design and reviewed drafts of the paper. }
\affiliation{Software Engineering Group, Institute of Software Technology, University of Stuttgart, Germany}



\paperdetails{
  perspective=scienceempirical,
  area={User Interface Programming},
  license=cc-by,
}



\paperdetails{
  submitted=2016-12-01,
  published=2017-04-01,
  year=2017,
  volume=1,
  issue=2,
  articlenumber=17,
}
\maketitle

\begin{abstract}
  \input{content/abstract.tex}
\end{abstract}

\input{content/content.tex}

\noindent 

\newpage

\printbibliography

\newpage
\appendix
\input{content/appendix.tex}

\end{document}

%% file: content/abstract.tex
The better developers can learn software tools, the faster they can start using them and the more efficiently they can later work with them. Tutorials are supposed to help here. While in the early days of computing, mostly text tutorials were available, nowadays software developers can choose among a huge number of tutorials for almost any popular software tool. However, only little research was conducted to understand how text tutorials differ from other tutorials, which tutorial types are preferred and, especially, which tutorial types yield the best learning experience in terms of efficiency and effectiveness, especially for programmers.

To evaluate these questions, we converted an existing video tutorial for a novel software tool into a content-equivalent text tutorial. We then conducted an experiment in three groups where 42 undergraduate students from a software engineering course were commissioned to operate the software tool after using a tutorial: the first group was provided only with the video tutorial, the second group only with the text tutorial and the third group with both. 

In this context, the differences in terms of efficiency were almost negligible: We could observe that participants using only the text tutorial completed the tutorial faster than the participants with the video tutorial. However, the participants using only the video tutorial applied the learned content faster, achieving roughly the same bottom line performance. We also found that if both tutorial types are offered, participants prefer video tutorials for learning new content but text tutorials for looking up \enquote{missed} information.

We mainly gathered our data through questionnaires and screen recordings and analyzed it with suitable statistical hypotheses tests. The data is available at~\cite{kulesz_daniel_2016_188896}. 

Since producing tutorials requires effort, knowing with which type of tutorial learnability can be increased to which extent has an immense practical relevance. We conclude that in contexts similar to ours, while it would be ideal if software tool makers would offer both tutorial types, it seems more efficient to produce only text tutorials instead of a passive video tutorial -- provided you manage to motivate your learners to use them.

%% file: content/content.tex
\section{Introduction}
Typical developers have to work with many different tools every day. While many developers get frustrated when the tools' user interfaces do not match their expectations ~\cite{lazar}, most developers manage to come to terms with what they get. However, as technology evolves, developers are expected to get along with new tools quickly. Thus, the learnability and understandability of a software tool is an important success factor~\cite{dix, grossman2}. Some software tool producers believe to tackle these issues appropriately by providing tutorials for the software. However, as Martin et al. have shown, tutorials are often unavailable, incomplete or focusing on the wrong aspects~\cite{martin}.

While in the early days of computing mostly text tutorials were available~\cite{texts}, advances in technology made it possible to easily produce video tutorials and even interactive tutorials. Van Loggem's research indicates that the \enquote{classic} written (and printed) manuals usually are not the first choice -- at least for most tool end-users today~\cite{loggem}. Instead, most tool users prefer interviewing colleagues or searching for tutorials in online sources, and especially the latter offer many different kinds of tutorials to choose from. Furthermore, many tutorials are not crafted by the original makers of the software or hired tutorial producers but by other tool users who post such tutorials on blogs or social media platforms.

In this study, we focused on passive tutorials, meaning tutorials which present information but do not react to the user in any way, as they are the most common ones for application software. We also used only text tutorials that were transformed from video tutorials.

\subsection{Problem Statement}
Our analysis of the existing literature showed that there are a few studies which address the differences between paper-based and video tutorials. The results are not yet conclusive, however. Some see advantages in video tutorials~\cite{baecker, vanderMeij, lloyd}, others in text tutorials~\cite{mestre} while also several studies found no difference~\cite{payne, alexander, deVaney}. None of these studies considered software developers and development tools in particular. Our focus his hence to understand which tutorial types are preferred by \textbf{developers} and which tutorial types yield the best learning experience for them in terms of efficiency and effectiveness. This is is a problem because developers are overburdened with too many choices while software tool makers do not know on which type of tutorials they should spend their (typically) limited resources.

\subsection{Research Objective}

Therefore our overall research objective is:\\
RO: What is the best way for developers to learn new software tools?

To address this research objective, we formulated the following research questions:

\begin{itemize}
	\item RQ1: What kind of tutorial do developers prefer if both text and video tutorials are available?
	\item RQ2: Which tutorial takes developers less time?
	\item RQ3: Which tutorial is more effective for developers?
\end{itemize}

\subsection{Context}

The study was conducted using 42 students from an undergraduate software engineering course. We used an experimental software for ing spreadsheets which is implemented as an add-in for Microsoft Excel~\cite{excel}. Since both the tool and its underlying approach are novel, learning and understanding them requires adequate tutorials. Therefore, results obtained from this study should be considered especially in contexts where developers are confronted with new software tools which are not directly mappable to previous experience. 

The study described in this paper borrows its design and task descriptions from another study conducted in early 2016 by the second author (the study has not been published yet). However, it is not a pure replication: while the original study only used video tutorials to explain SIF, this study also used content-equivalent text tutorials which were produced for this very study. Furthermore, the participants were asked more questions about their perceptions on the tutorials they consumed, while this aspect was not targeted in-depth in the original study.

This paper follows the structure proposed by Jedlitschka et al.~\cite{jedlitschka} with a slight deviation (we describe the procedure of the experiment earlier than proposed by Jedlitschka et al. and merged the alternative technologies and the related studies): After describing the background of the study, we explain the plan of the experiment and its deviations, we present the obtained results and discuss them before drawing a final conclusion and outlining future work. The graphical representations are based on the recommendations by Tufte~\cite{tufte}.

\section{Background} 
\subsection{Technology under Investigation}
\label{subsec:technology}

The \textit{Spreadsheet Inspection Framework} (SIF)~\cite{sifei} is a software application for  faults in spreadsheets which has been developed at the University of Stuttgart and already has been evaluated in a number of previous studies~\cite{kulesz1,kulesz2,kulesz3}. It is operated through a plug-in for Microsoft Excel which allows end-users to automatically scan spreadsheets for \enquote{bad smells} (e.g.\ formulas referencing the same cell twice) or violations of design rules (e.g.\ formulas with constants). Apart from running these pre-defined scans, it also allows end-users to specify their own test scenarios which are comparable to unit tests for normal software applications. While SIF is primarily targeted at end-users, we expected developers to understand the tool's concepts faster, because these concepts are similar to static analysis and unit tests in traditional software development.

\begin{figure}
	\includegraphics[width=\textwidth]{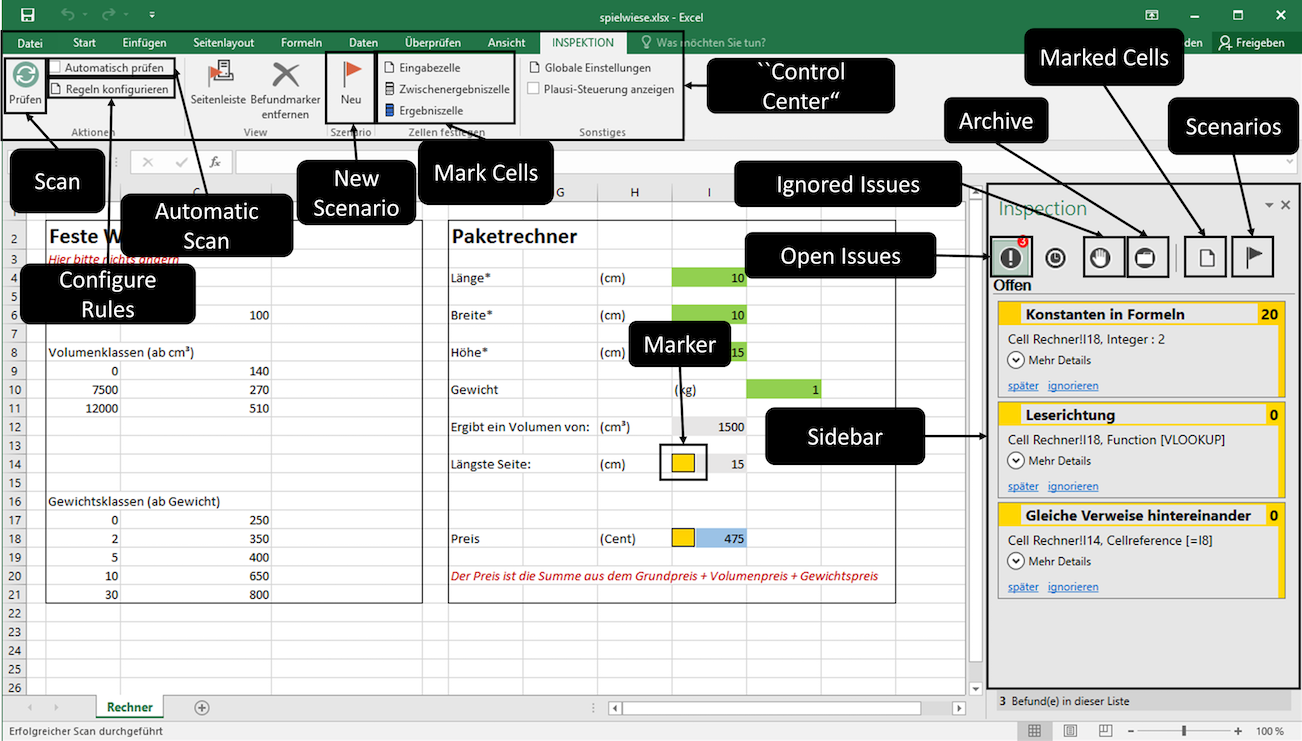}
	\caption{Main user interface of the Spreadsheet Inspection Framework}
	\label{fig:sif}
\end{figure}

A screenshot of the main user interface of SIF is provided in Fig. \ref{fig:sif}. It shows that SIF adds a new tab on the ribbon bar which acts as SIF's \enquote{control center}. Here, the user can configure the rules to be used for testing spreadsheets, create new test cases and start the automated scans of the spreadsheet. The sidebar on the right side provides the user with an overview about open, postponed, ignored and solved findings. Furthermore, the issues reported in the sidebar are synchronized with marker symbols in the corresponding cells of the spreadsheet. Although none of these elements of the user interface is overly complex, learning how to use them all in just a few minutes can be challenging for end-users -- making this setting a proper ground for evaluating the efficiency and effectiveness of different tutorials.

For the purpose of another study~\cite{other_study}, the second author has produced an introduction video and two video tutorials (educational screencasts) which explain the ideas behind SIF and how to use it. Since we regard comparing tutorials with different (depth of) content as not being fair, the first author simply converted the video tutorials into content-equivalent text tutorials, making them directly comparable. This also ensured that the videos had the same origin and thus were comparable to \enquote{official tutorials} produced by an original software maker (unlike tutorials made by hired tutorial producers or third parties).

\subsection{Related Studies} 
\label{subsec:related}

Overall, there are only a few studies which investigated tutorials for software. From these studies, we do not have a conclusive understanding.
On the one hand, a study by Mestre~\cite{mestre} indicated that learners favored a static text tutorial over video tutorials.
On the other hand, several studies showed that learners prefer video tutorials over text tutorials~\cite{baecker, vanderMeij, lloyd}. 
On top of that, a study by Palmiter and Elkerton~\cite{palmiter} came to the conclusion that both text tutorials and video tutorials have their specific benefits.
And there are even studies that could not find significant differences between text and video tutorials at all (e.g.~\cite{payne, alexander, deVaney}). 

Our impression from these studies is that both video and text tutorials have their particular advantages and disadvantages. For video tutorials Alexander reported that learners watch the whole video(s) (which is time consuming)~\cite{alexander} instead of picking particular scenes. However, studies by Mestre and Alexander both confirm that learners have trouble finding specific scenes after watching a video~\cite{mestre, alexander}. As an advantage of video tutorials, participants in a study by deVaney perceived them as \enquote{more human}~\cite{deVaney}.  For text tutorials, the study by Mestre came to conclusion that two huge advantages are that (i) the participants can decide which part of it they want to read and (ii) that information in a text can be found faster than in a video~\cite{mestre}.

Overall, it was not possible to find a common denominator regarding to which tutorial type is the best one. Additionally, none of the mentioned studies investigated programmers in particular. The aim of our study was to fill this gap.

\subsection{Practical Relevance}
The better users can learn software tools, the faster they can start actually using it and the more efficiently they can later work with it. Thus, good learnability is a vital success factor for software tools. Therefore, knowing with which type of tutorial learnability can be increased has an immense practical relevance. 

Furthermore, producing tutorials also requires effort. In our own experience, we perceived producing video tutorials to be far more difficult than producing text tutorials (the second author worked about 80 hours to produce three video tutorials with a total length of less than 25 minutes). Therefore, knowing if video tutorials actually have advantages in terms of learnability is also relevant for practice.

\section{Experiment Planning}

\subsection{Goals} 
For every research question we defined a goal:
\begin{itemize}
	\item Goal 1: Analyze video and text tutorials \\
	For the purpose of understanding which tutorial type is preferred\\
	With respect to the preference rate of the users for each type
	\item Goal 2: Analyze video and text tutorials\\
	For the purpose of comparing their efficiency\\
	With respect to the time required by users to complete the tutorials and the tutorials plus a following  tasks
	\item Goal 3: Analyze the quality of video and text tutorials\\
	For the purpose of comparing their effectiveness\\
	With respect to counting how often users looked something up, the number of wrong actions they took, measuring the perceived difficulty level of the tutorials and with respect to which percentage of the learning content they understood right away	
\end{itemize}

~

Depending on the context, the meanings of the terms efficiency and effectiveness can vary. Therefore, we want to give our definition of these terms in the context of this paper. An explaination how we measured them will be presented in section \ref{sec:analysis_procedure}.

\paragraph{Efficiency} Efficiency is the ratio between input and output. In our case, we investigated the ratio of the content of the tutorials to the time needed to consume this content. The faster a tutorial teaches some content, the more efficient it is.

\paragraph{Effectiveness} Effectiveness measures to what degree a goal was reached with a given method. In our case, there are two types of effectiveness: (I) effectiveness in terms of using a tutorial as a source to look up information and (II)  effectiveness in terms of using a tutorial to learn how to solve a task. The easier information can be looked up in a tutorial and the better it teaches how to solve a task, the more effective it is.

\subsection{Participants}
To successfully pass one of the undergraduate software engineering lectures, all students were required to take part in an experiment. We decided to use this population primarily because it was easily accessible. There were 42 participants and they were about 20 years old. Nine of them were female and thirty-three were male.

We gave the participants the choice between 14 time slots for doing the experiment, reserving them on a \enquote{first-come-first-serve} basis. We ran three parallel experiments in each time slot, strictly separated as described in section \ref{subsec:procedure}.

Each participant had to fill out an online questionnaire a few days before the experiment. Based on the completion order of this questionnaire we assigned each participant a unique anonymous ID.

To ensure confidentiality, each participant had to sign an agreement at the beginning of the experiment about not disclosing any details of the experiment to other participants. Additionally, we stressed this point at the beginning of each experiment and explained the motivation behind this. All participants were allowed to abort the experiment at any time if they wished but no participant used this opportunity. 

\begin{figure}[b]
	\includegraphics[clip, trim=2.5cm 0.5cm 2.5cm 0.5cm,width=\textwidth]{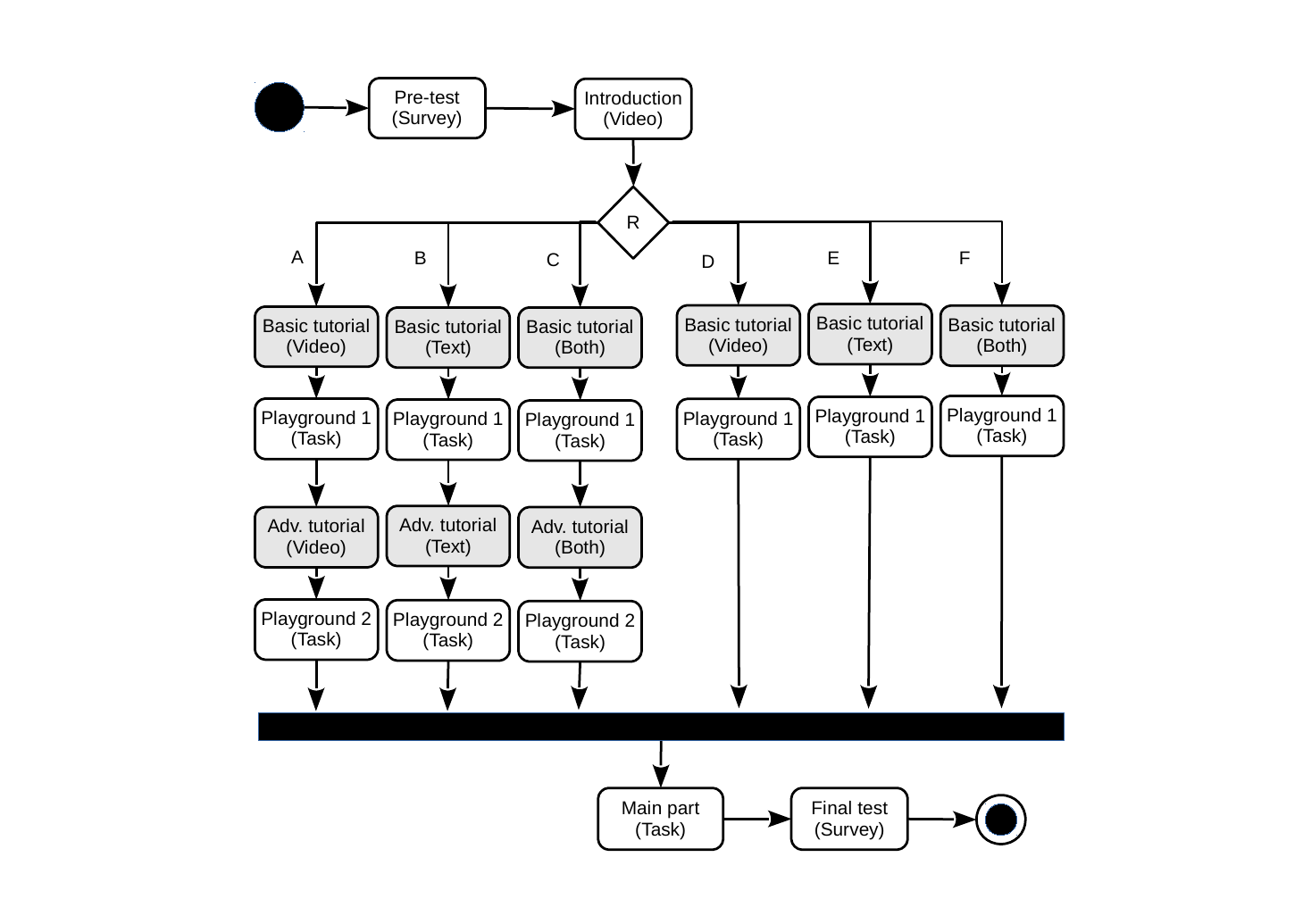}
	\caption{Overview of the experiment procedure (slightly simplified)}
	\label{fig:evalconcept}
\end{figure}

\subsection{Procedure}
\label{subsec:procedure}
Figure \ref{fig:evalconcept} provides an overview of the procedure in our study. The actual steps where participants use tutorials are highlighted in light gray. However, we simplified the \enquote{final test} in this figure to make it more concise -- in reality, some parts of the questionnaire were dependent on the type of experiment the participant had.

As it can be seen, we randomly divided the participants into six groups. Every group contained seven participants.

\begin{itemize}
	\item A: with scenarios, video only
	\item B: with scenarios, text only
	\item C: with scenarios, video and text
	\item D: no scenarios, video only
	\item E: no scenarios, text only
	\item E: no scenarios, video and text
\end{itemize}

All participants had to watch the introduction video. After using their particular tutorial, all participants solved the first task (Playground 1). After finishing the first task, the participants of groups A, B, and C directly continued with the main task. By contrast, participants of the other groups first had to do another round of tutorials in which they learned the (advanced) scenario testing technique and solved simple tasks with it (Playground 2) before continuing with the main task. 

Since each of our time slots had the capacity for three participants, we either executed experiments A, B and C or D, E and F (the idea behind this was not to give participants the feeling that they are slower than others if they see that others who started at the same time had already finished). The particular assignment to A, B or C or D, E or F respectively was based on the order of arrival of the participants.

Before the actual start of the experiments, we explained to each group of three participants the whole setup and what they had roughly to do, not giving them any details except \enquote{you will get instructions where everything will be explained}. The experiments took place in one of our computer pools with air-conditioning, on standard desktop computers with 23" monitors with a full HD resolution. The \enquote{work-places} of the participants were close to each other but we put separating walls in-between as shown in figure \ref{fig:pool} (appendix), so the participants could only see their individual screen.

For later examination, we recorded all experiments using a screencast recorder. Also, we watched the participants' screens over VNC connections during the experiment thus avoiding to directly look over their shoulders so as to not make them feel uncomfortable. This was a tough trade-off, as it also made it impossible for us to count how many times the participants looked something up in the text tutorials ourselves.

There were some common issues which most participants encountered, such as not knowing how the VLOOKUP formula works in Microsoft Excel. In such cases we gave the participants \enquote{meta hints} like "maybe you could try to look this up on the internet?". We allowed access to the internet because it is a very common way to retrieve information about unknown formulas.

SIF also has several known bugs and we were aware of them already before the experiments (they are not straightforward to fix). If the participants encountered such a bug we helped them get around it. We took extreme care not to disturb the other participants in such cases as we did not want to influence them.

\subsection{Experimental Material}
The participants of our experiment had to learn how to use the \textit{Spreadsheet Inspection Framework} (SIF). 
The introduction video had a length of 09:03 minutes. The video about the basics of SIF was 05:42 minutes long and the scenario video had a length of 10:01 minutes. The introduction video explained general problems of spreadsheets and the concept of design patterns for spreadsheets, whereas the other two videos explained the tool SIF itself (what does it do), how to use static rules and how to create a scenario. The videos were not chunked or split in chapters. They did not cover all functions of SIF but all functions that were necessary to solve the given tasks. 

Following existing studies~\cite{plaisant, mestre, lloyd, howto}, we decided to have videos with a screen recording and narration.
In the videos we use two different techniques similar to the ones used by Ilioudi et al.~\cite{videos}. Either the left side of the screen showed a screen-cast of a spreadsheet being manipulated and the right side showed the second author explaining what he does in the spreadsheet (Talking Head Style) or the whole screen showed the screen-cast of a large spreadsheet with an explaining voice in the background (Khan Style). In every case the clicked buttons and important parts of the GUI were highlighted with blue borders and arrows. An example can be seen in Fig.~\ref{fig:video} in appendix \ref{app:tut}. The participants were in full control of the video, meaning they could use every  option given by the Windows Media Player.

The text tutorials were directly generated from the video tutorials, thus following the same explanatory scheme. 
Still, we decided to rephrase some sections which were quite colloquial in the videos. 
The text tutorial counterpart for this excerpt of the video is illustrated in Fig.~\ref{fig:leiste} in appendix \ref{app:tut}. As it can be seen, similar highlighting techniques have been used in the text to explain which actions the user has to take to reach a certain functionality.

In the pre-test, we asked several questions using an online questionnaire to investigate the participants' experience and opinions. The pre-test was made up of a questionnaire which contained questions about prior knowledge of Microsoft Excel and what tasks the users typically solve using it, if any. It can be seen in appendix \ref{app:pretest}. 

During the experiment, each participant received printed instructions on what to do next (obviously, the instructions differed depending on which experiment the participant had). The instructions also contained questions about the last task they completed, so the participants worked their way through the instructions, alternating between reading the instructions, doing activities on the computer and writing down answers on the instruction sheets. An example of the instructions is shown in Fig. \ref{fig:instr} in the appendix.

Apart from the instructions, we also provided the participants with the (printed) text tutorials, the tutorial videos or both (depending on the group). After finishing all practical tasks, each participant was asked to fill out a paper questionnaire. Here, we asked the participants to self-estimate what percentage of their tutorial they understood when they consumed it for the first time. We also asked them as how difficult they perceived the whole study. Finally, we asked more questions about their background, what they liked and disliked about SIF and, of course, about the tutorials. An example of the final questionnaire for group C can be seen in appendix \ref{app:final_questions}.

Last but not least the participants were provided with either two or three spreadsheets (depending on the group) which contained a number of seeded errors to solve the actual tasks.

\subsection{Tasks} 
\label{subsec:tasks}
Depending on the group of the participants they had to perform either two or three bigger tasks during the experiment. These bigger tasks were split up into a series of small sub-tasks.

The first bigger task (Playground 1) was to apply the basic functions of the SIF. First, the participants were asked to open a spreadsheet, to activate several static rules and to initiate a scan. For this we prepared a spreadsheet that could be used to calculate the price of a parcel based on its size and weight. If done right, SIF reported for this spreadsheet an issue with the reading direction which could be solved by moving the content in one of the cells. The second issue reported by SIF was that one formula referred to the same cell multiple times using the MAX function (The formula was: MAX(I4;I6;I8;I8;I8;I8;I8;I8;I8). This issue could be solved by removing the obsolete references (it makes no sense to refer to the same cell multiple times in a formula which is composed only of the MAX function).

The second task (Playground 2) was only performed by participants in the groups that had received the advanced training where they were taught how to use the SIF scenario testing technique. The participants were first asked to create a new scenario on their own and then to use it to find errors in a new spreadsheet. Again we used the example of the parcel price. The created scenario reported an issue because for the final price calculation the weight of the parcel was subtracted. The participants solved this by repairing the formula. 

The last and lengthiest task was to extend a given spreadsheet by adding new data and features. The given spreadsheet calculated the monthly bill for using a mobile phone based on a user's consumption of minutes and text messages (with different rates based on the destination) for different tariffs (which had different rates for these minutes and text messages). It also featured a dashboard where monthly bills could be compared between the tariffs to find the cheapest one. The participants first had to add a new rate and then add the costs for on-net texts for every rate.

\subsection{Hypotheses}
We tested the following hypotheses. All hypotheses have only the scope of developers learning a software tool from
a static text tutorial, a non-interactive video tutorial or both. The numbers refer to the research questions. 
\begin{itemize}

	\item ${H_1}$: The video tutorial is chosen more or less often than the text tutorial when both are provided.

	\item ${H_{2a}}$: The three tutorial groups differ in the time they need to complete the tutorials.

	\item ${H_{2b}}$: The three tutorial groups differ in the time needed to complete the tutorials and the following tasks.

	\item ${H_{3a}}$: The participants look up more or less items in the video tutorial than in the text tutorial.

	\item ${H_{3b}}$: The participants differ in their understanding of the video and the text tutorial at the first attempt. 

	\item ${H_{3c}}$: The three tutorial groups differ in the number of wrong answers given during the tasks. 

	\item ${H_{3d}}$: There is a difference in how difficult the participants thought the tutorial was.
	
\end{itemize}

\subsection{Analysis Procedure}
\label{sec:analysis_procedure}

To analyze the results, we examined the screen recordings and the questionnaires. In the screen recordings, we measured when a participant started a task (when he or she opened the Excel file) and when the participant finished the task (closing the file). The time between one file and the next one was the time used for the tutorial. We measured the duration in seconds and combined the groups with the same used tutorial for the SIF part.

For evaluating the correctness of the results, we designed unit tests with input and expected output values -- one unit test for each playground and two unit tests for the final task. We had two unit tests for the final task because the final task included two tasks -- the the new rate and then the on-net texts. We then filled in the input values and checked if the actual output values matched the expected values. We judged a unit test for a spreadsheet to be correct if and only if \emph{all} actual output values matched their expected values.

For the questionnaires, we first used LimeSurvey~\cite{limesurvey} to produce an electronic version of the questionnaires. Then we exported them to R~\cite{r} where we evaluated them. 

We first determined with a Shapiro-Wilk test if the samples were normally distributed. As the sample groups were quite small, we then used either a t-test t (normally distributed) or a Wilcoxon signed-rank test w (otherwise) to see if the group results were significantly different. For this, we chose a significance level of 0.05. Additionally, we measured the effect size with Cohen's d and the absolute mean difference MD. 
To check the normal distribution in section \ref{H1} we used a Shapiro-Wilk test due to a small sample size with a significance level of 0.05. 

To measure efficiency, we analyzed the time the participants needed to consume the tutorials. Independent from that, we analyzed the time the participants needed to complete the corresponding playground tasks. We decided to take these separate measures because we wanted to investigate the time distribution between consuming the tutorials and solving the tasks.

To measure effectiveness, we had to consider both types of effectiveness. For the first effectiveness type (looking up information) we counted how often the participants looked something up in the tutorials to see which tutorial is more useful for this task. We would rate the tutorial which was used more often as more effective. For the second effectiveness type (learning how to solve a task) we analyzed the participants' estimates regarding their understanding of the tutorials and their perceived difficulty of the study. Apart from these subjective measures, we also counted the number of wrong actions they took -- the less wrong actions they took, the better.

\section{Execution} 
\subsection{Preparation} 
Before the participants arrived, we prepared the computers and started the screen broadcast over the VNC connection. For the participants, there was no particular preparation as all required information was provided during the experiment via the instruction sheets and the particular tutorial. Therefore, we did not ask the participants to prepare themselves in advance.

\subsection{Deviations} 
\label{sibsec:deviations}
There were only two deviations from the plan:

\begin{itemize}
\item By mistake, we used one of the computer accounts twice, so that the data of the previous participant was overwritten. However, thanks to the screen recording we were able to redo every single action taken by this user in the experiment, recreating the lost spreadsheet.
\item One of the participants had issues with his e-mail account so he did not get our invitation for his particular time slot. Therefore, we ran only the remaining two experiments in this time slot. Then, we simply added a new time slot where this participant did the missed experiment alone.
\end{itemize}

\section{Analysis} 
\subsection{Descriptive Statistics}

\subsubsection{Goal 1 -- Usage of the Tutorials}
\begin{figure}
	\centering
	\includegraphics[width=0.9\linewidth]{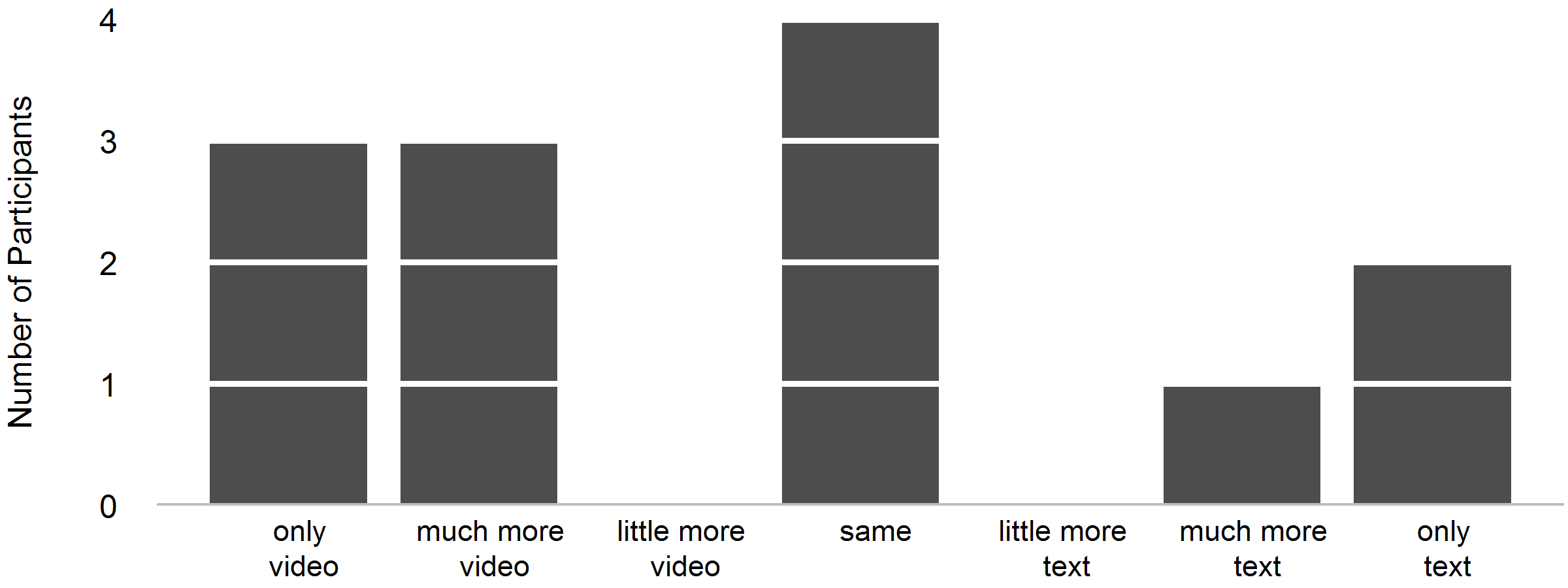}
	\caption{Question for the participants with both tutorials: \enquote{which tutorial did you use more often?}}
	\label{fig:moreOften}
\end{figure}

For this goal we asked the groups with both tutorials which tutorial they used more often. Fig.~\ref{fig:moreOften} shows the results. As can be seen, from the 13 participants in group C and F who answered this question, six participants used (nearly) only the video tutorial, four used both equally and only three used (nearly) only the text tutorial.

\subsubsection{Goal 2 -- Duration of Tutorial Usage}
\begin{figure}
 	\centering
 	\subcaptionbox{Duration of using the SIF tutorial in seconds\label{fig:duration_T1}}%
 	[.49\linewidth]{\includegraphics[width=0.49\linewidth]{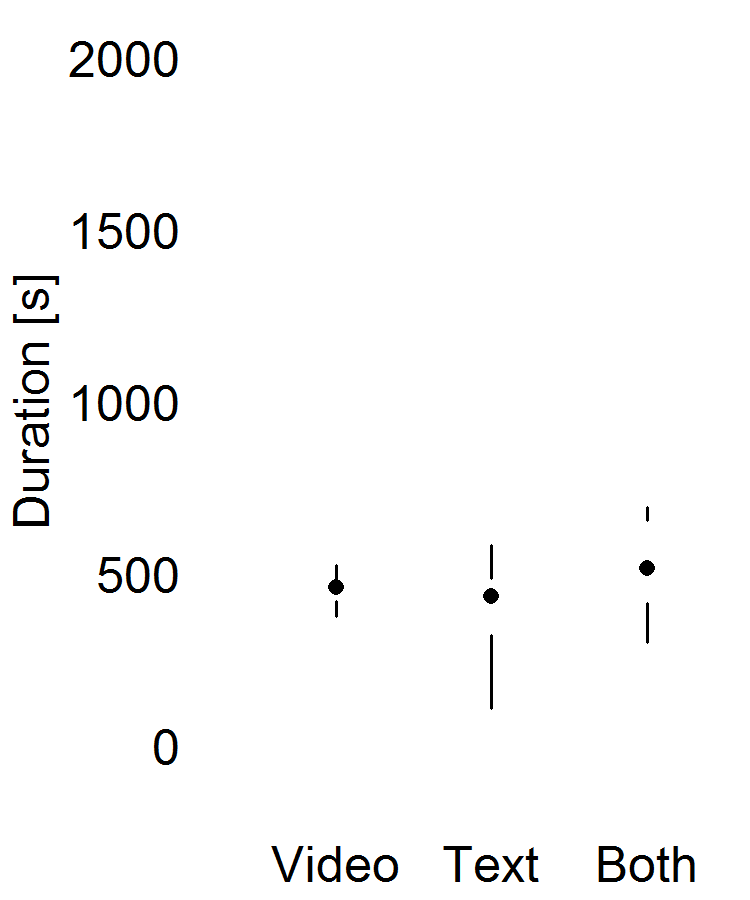}}
 	\subcaptionbox{Duration of using the SIF tutorial and the following task in seconds\label{fig:duration_SW1TUT}}
 	[.49\linewidth]{\includegraphics[width=0.49\linewidth]{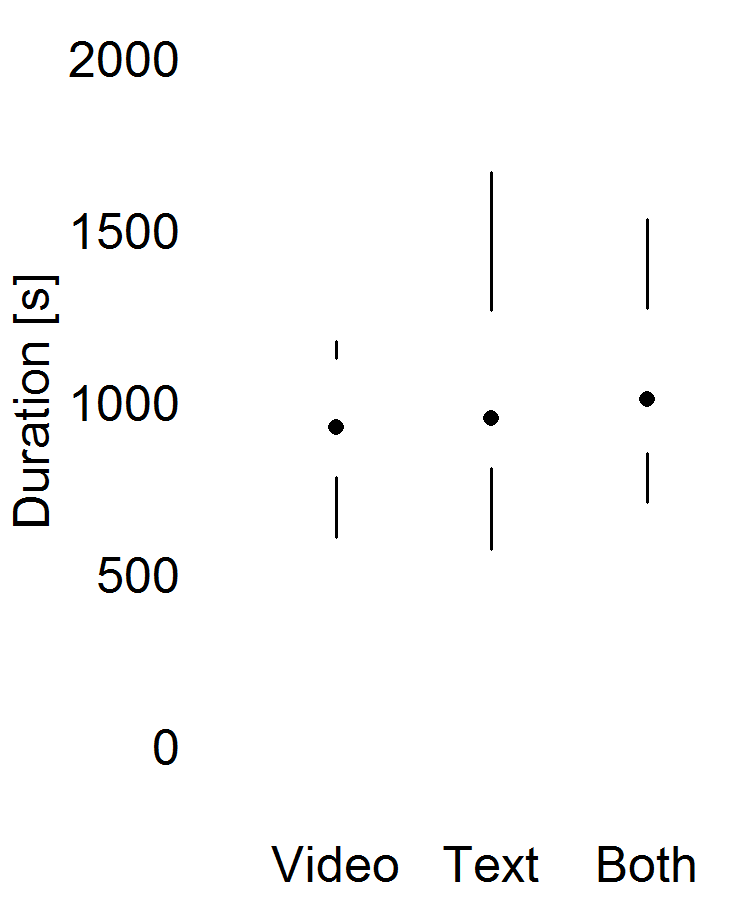}}
 	\caption{Duration of using the SIF tutorial and using the SIF tutorial plus the following task in seconds}
 \end{figure}

 Fig.~\ref{fig:duration_T1} shows that for the SIF tutorial there were only small differences between the time needed by the video and the text tutorial. The participants with both tutorials took more time to complete the tutorial. The standard deviation of all three groups is close around the mean with the deviation of the \textit{video and text} group slightly larger. ($\sigma_{\mathit{Video}}=117.361$, $\sigma_{\mathit{Text}}=139.764$, $\sigma_{\mathit{Both}}=188.432$).
 
  Furthermore, it can be seen in Fig.~\ref{fig:duration_SW1TUT} that there were only small differences in the overall time needed for the tutorial and the following task. The values spread wider around the mean ($\sigma_{\mathit{Video}}=362.353$, $\sigma_{\mathit{Text}}= 450.317$, $\sigma_{\mathit{Both}}=344.713$).
 
\begin{figure}
 	\centering
 	\subcaptionbox{Duration of using the scenario tutorial in seconds\label{fig:duration_T2}}%
 	[.49\linewidth]{\includegraphics[width=0.49\linewidth]{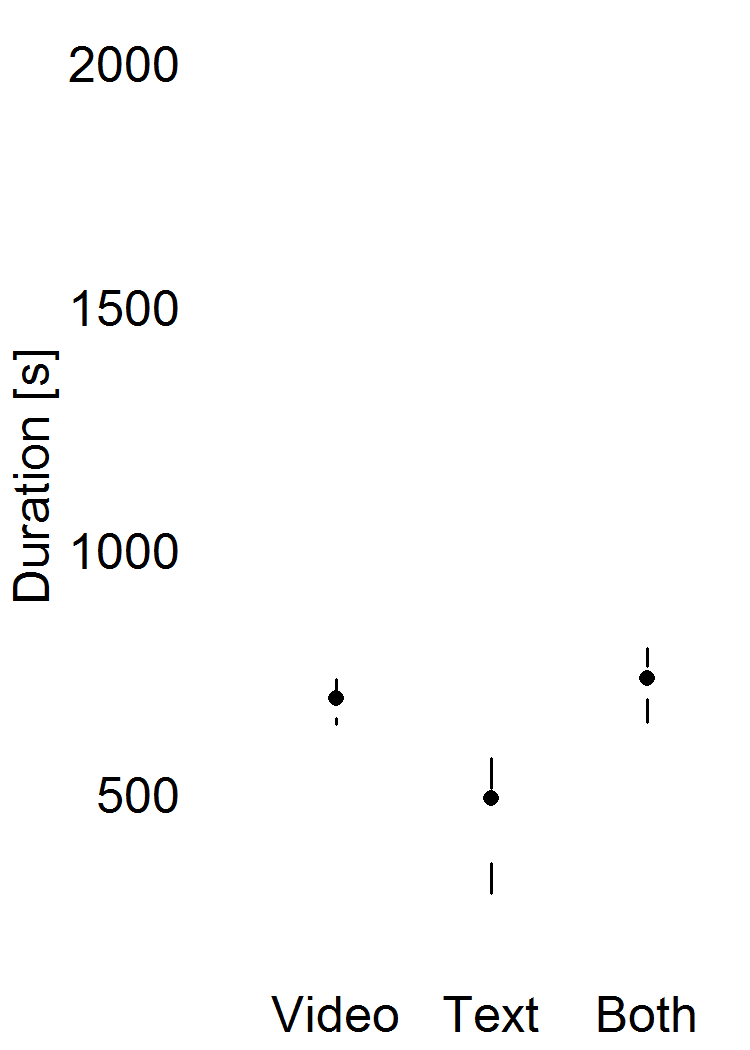}}
 	\subcaptionbox{Duration of using the scenario tutorial and the following task in seconds\label{fig:duration_SW2TUT}}
 	[.49\linewidth]{\includegraphics[width=0.49\linewidth]{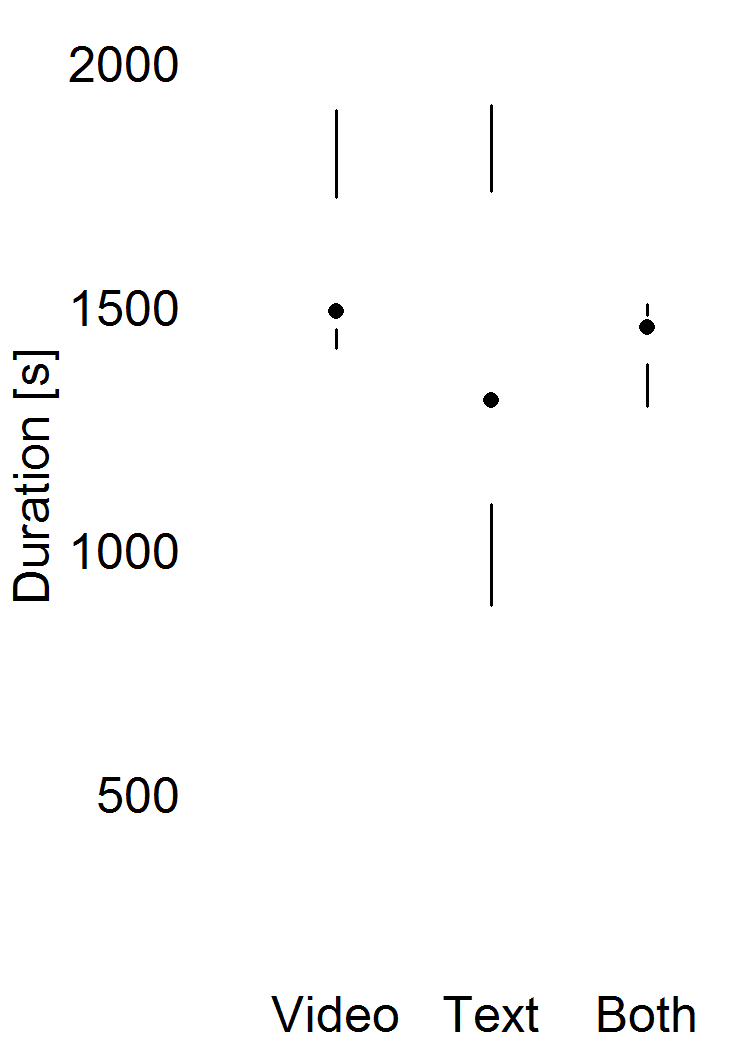}}
 	\caption{Duration of using the scenario tutorial and using the scenario tutorial plus the following task in seconds}
 \end{figure}

For the second tutorial the results were slightly different. Fig. \ref{fig:duration_T2} shows that this time the text tutorial was much shorter than the other two tutorial types. The values are much closer around the mean than for the SIF tutorial ($\sigma_{\mathit{Video}}=362.353$, $\sigma_{\mathit{Text}}= 103.714$, $\sigma_{\mathit{Both}}=54.705$).
But again there are close to no differences when comparing the tutorial and the following tasks as can be seen in Fig.
 \ref{fig:duration_SW2TUT}. Again, the values are further away from the mean ($\sigma_{\mathit{Video}}=396.333$, $\sigma_{\mathit{Text}}= 357.612$, $\sigma_{\mathit{Both}}=126.839$).

\subsubsection{Goal 3 -- Effectiveness of the Tutorials}

\begin{figure}
	\centering
	\subcaptionbox{The participants with \textbf{only one} tutorial stated how many times they looked something up in the used tutorial\label{fig:lookingUp}}%
	[\linewidth]{\includegraphics[width=\linewidth]{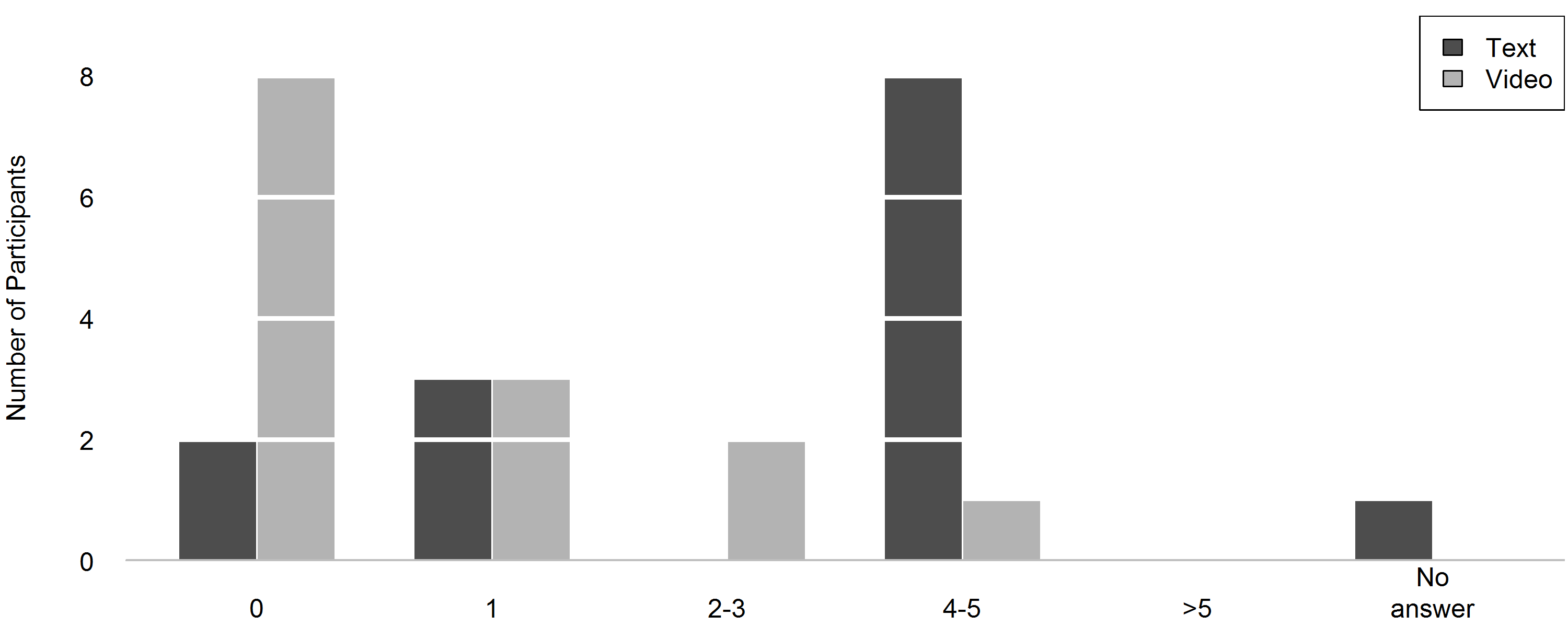}}
	\subcaptionbox{The participants with \textbf{both} tutorials stated how many times they looked something up in which tutorial\label{fig:lookedUpBoth}}
	[\linewidth]{\includegraphics[width=\linewidth]{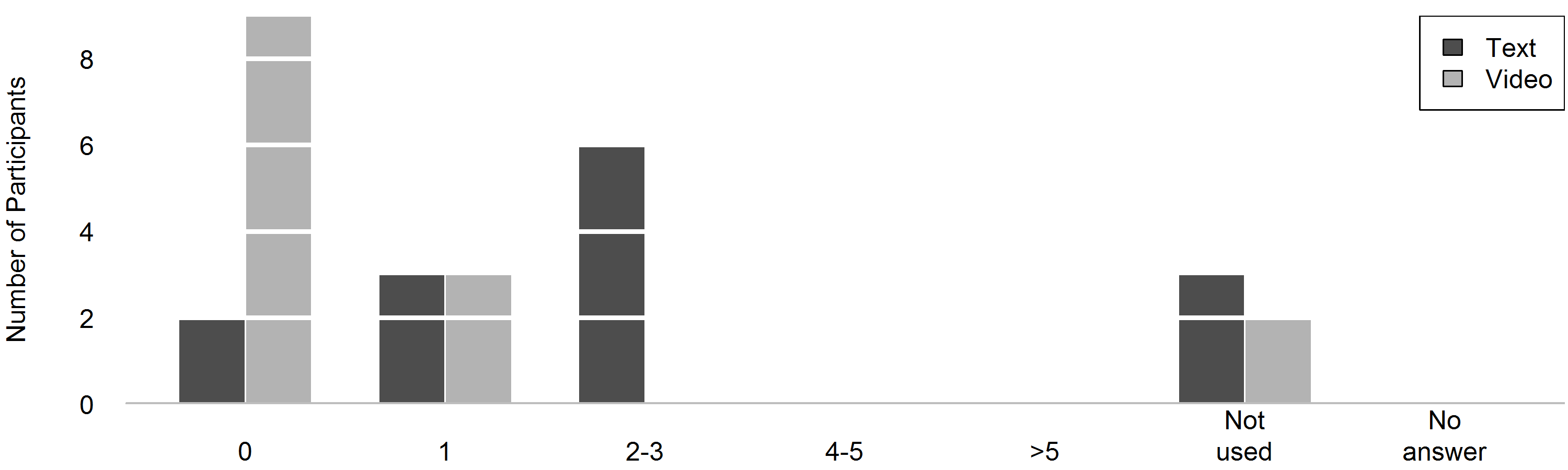}}
	\caption{How many times was an item looked up in which tutorial?}
\end{figure}

Fig. \ref{fig:lookingUp} shows how many times something was looked up in a tutorial by the two groups with text \textbf{or} video tutorial. The figure shows that the participants seldom looked something up in the video tutorial whereas they looked up things many times in the text tutorial. 

Fig. \ref{fig:lookedUpBoth} shows how many times the participants looked something up if they had \textbf{both} tutorials. Some participants used only one of the two tutorials (the other one was \textit{not used}). Again, they looked up more things in the text tutorial.

\begin{figure}
	\centering
	\includegraphics[width=0.5\linewidth]{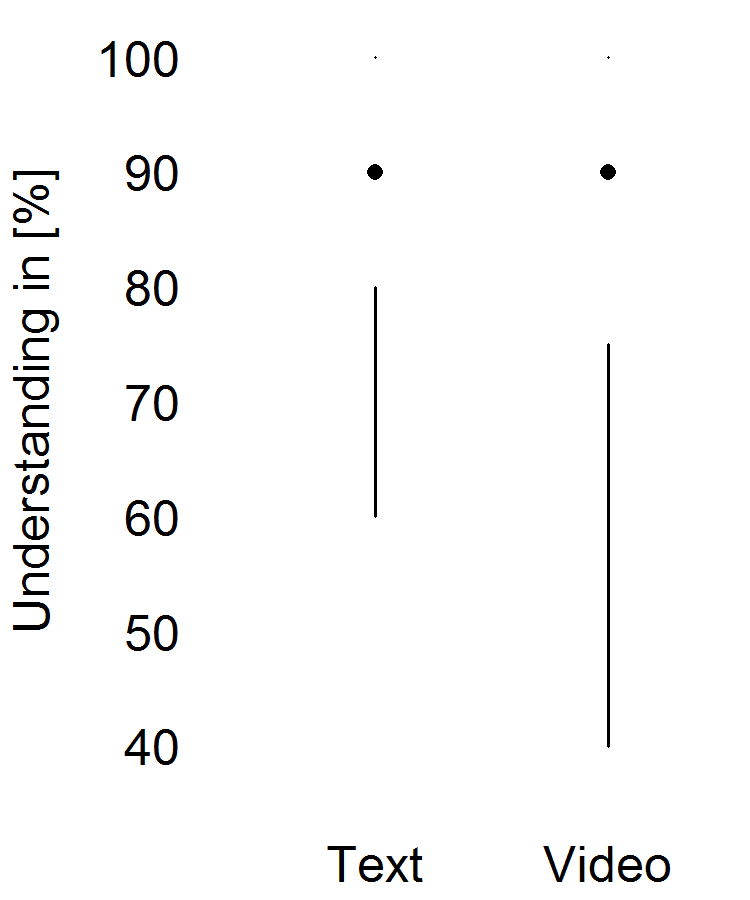}
	\caption{Q: What percentage of the tutorial did you understand when reading/watching it for the first time?}
		\label{fig:understandBox}
\end{figure}

Fig. \ref{fig:understandBox} shows how much the participants understood of the tutorials at the first attempt as assessed by themselves. In this case, the box plots have no upper whiskers and the little dots indicate the ends of the upper quartiles. There is nearly no difference between the text and the video tutorial. The values spread close to the mean ($\sigma_{\mathit{Video}}=18.321$, $\sigma_{\mathit{Text}}= 11.095$).

\begin{figure}
	\centering
	\subcaptionbox{Number of wrong and correct answers in the first task\label{fig:SW1}}%
	[0.45\linewidth]{\includegraphics[width=0.45\linewidth]{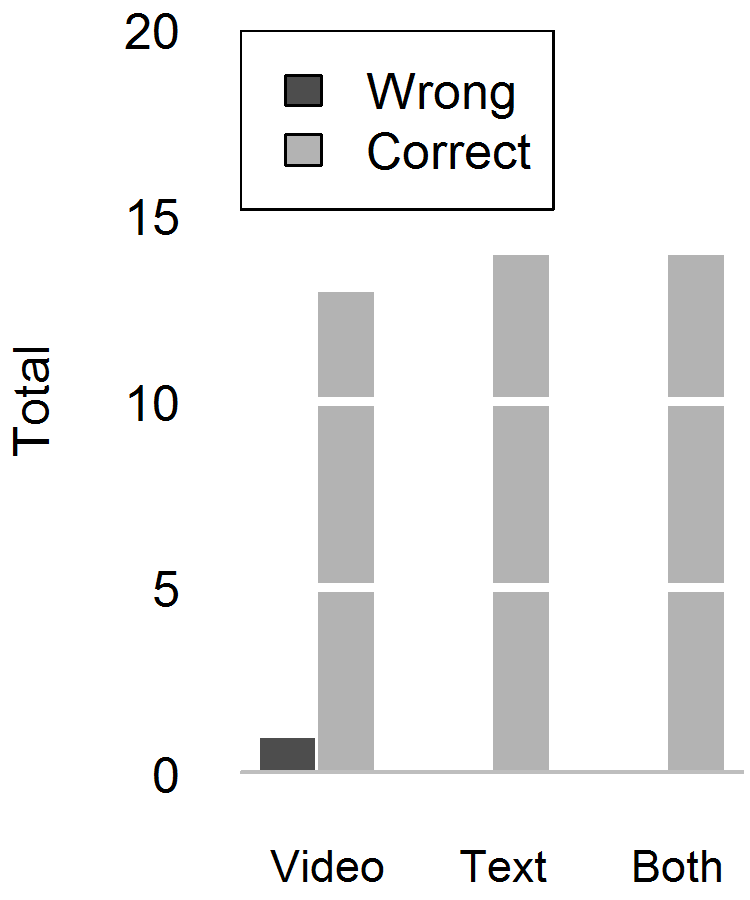}}
	\subcaptionbox{Number of wrong and correct answers in the second task\label{fig:SW2}}
	[0.45\linewidth]{\includegraphics[width=0.45\linewidth]{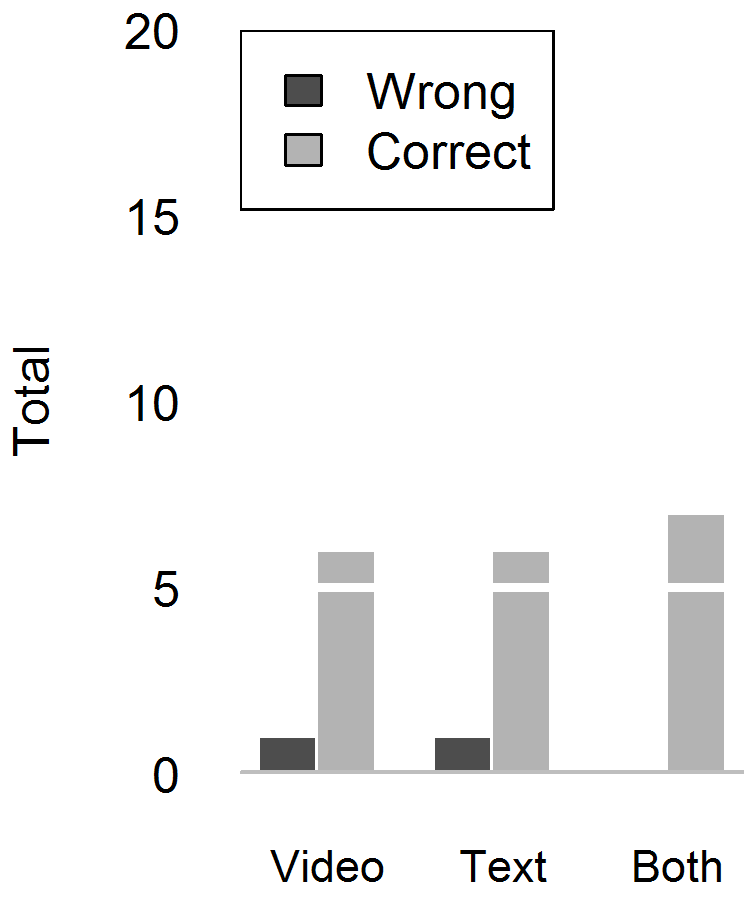}}
	\subcaptionbox{Number of wrong and correct answers in the final task after inserting the new rate\label{fig:finala}}
	[0.45\linewidth]{\includegraphics[width=0.45\linewidth]{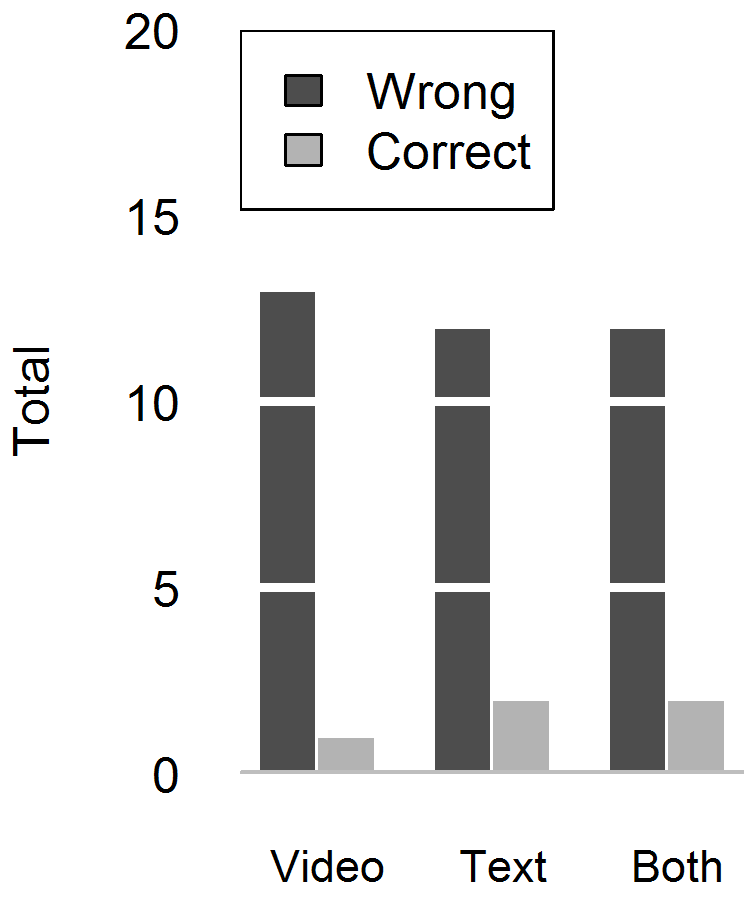}}
	\subcaptionbox{Number of wrong and correct answers in the final task after inserting the on-net texts\label{fig:finalb}}
	[0.45\linewidth]{\includegraphics[width=0.45\linewidth]{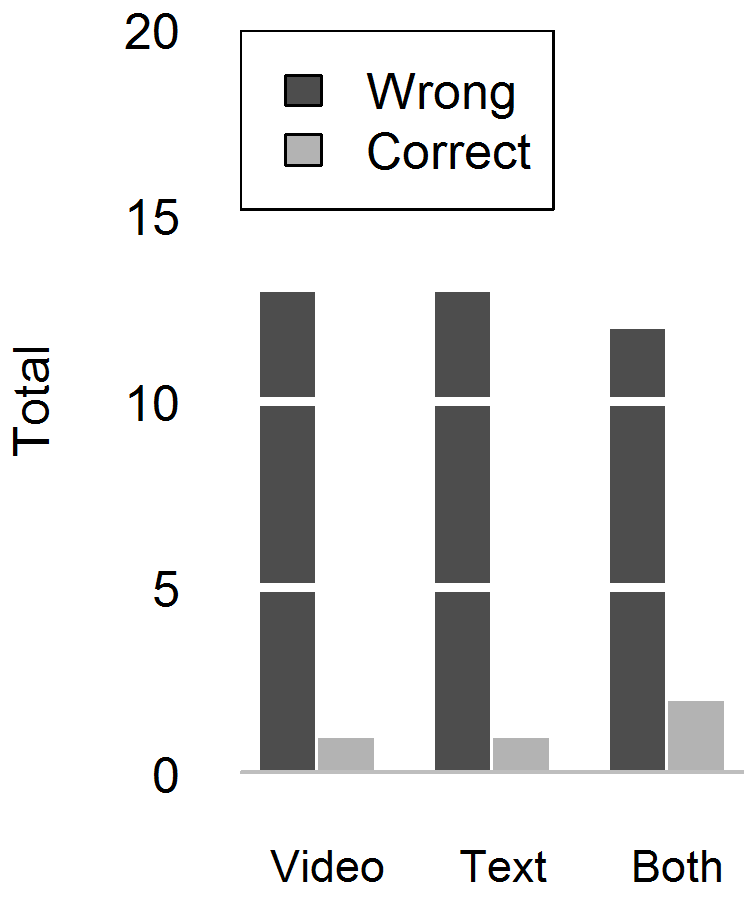}}
	\caption{Number of wrong and correct answers in the tasks}
	\label{fig:wrong_answers}
\end{figure}

In Fig.~\ref{fig:wrong_answers} it can be seen how many correct and wrong answers the result files of the three tasks contained. The files from the first two tasks (Playground1 and Playground2) contained almost no errors. Conversely, the correctness of the files in the main task was very poor. As mentioned, we evaluated two unit tests for this task. As Fig. \ref{fig:finala} shows, there were no significant differences between the three groups regarding the correctness of the first final task unit test. Also, when taking into account the second final task unit test the overall picture remains unchanged as shown in Fig. \ref{fig:finalb}.

\begin{figure}
	\centering
	\includegraphics[width=\linewidth]{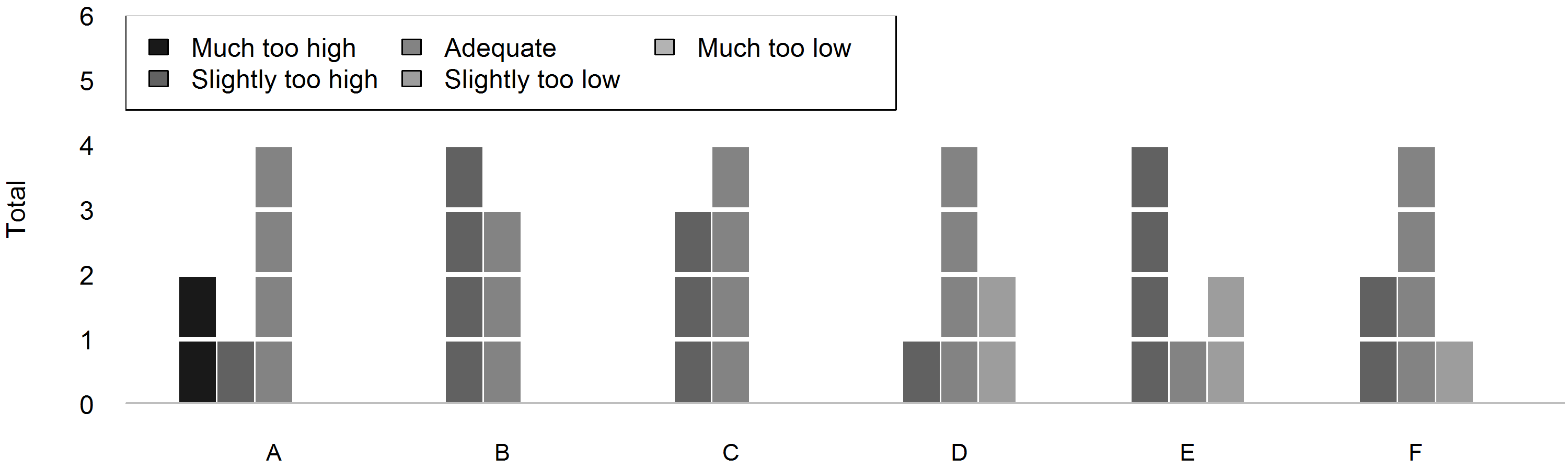}
	\caption{Difficulty level of our study perceived by the participants}
	\label{fig:niveau}
\end{figure}

Finally, Fig. \ref{fig:niveau} shows that there are small differences between the groups in how easy or hard they perceived the whole experiment to be. There is no clear tendency that one group had more or less challenges when solving the tasks in our study.

\subsection{Data Set Preparation}
To reconstruct the overwritten data mentioned in section \ref{sibsec:deviations}, one of us watched the screen recording and in parallel repeated the clicks from the video. 

Depending on the examined variables, we combined the results of some groups. For example, for the question which tutorial was used, we merged the two groups with both tutorials (C and F), no matter if with or without the scenarios as it was irrelevant in this case. 

One problem was the question for groups C and F how many times they looked up information. Many used the option \textit{Not used} as used zero times and not as the intended \textit{I did not use this tutorial at all}. We found out about this because many participants stated that they used both tutorials equally in one question and then stated that they did not use one at all in the described question. We solved this problem by correcting the \textit{Not used} answers to \textit{$0$ times} when the participants did not state \textit{only video} or \textit{only text} in the other question.  

\subsection{Hypothesis Testing}

\paragraph{$\mathbf{H_1}$ -- Usage of the Tutorials}
\label{H1}
The bar chart in Fig. \ref{fig:moreOften} looks like there might be a tendency towards the video tutorial. We conducted a Shapiro-Wilk test and saw that the data is normally distributed, but there is no preference for videos ((t) p=0.065). We cannot reject the null hypothesis.

\paragraph{$\mathbf{H_{2a}}$ -- Duration of Tutorials}
As Fig. \ref{fig:duration_T1} shows, the \textit{video and text} tutorial group took longer than the other two groups to complete the SIF tutorial.
The text tutorial group was indeed significantly faster than the slowest \textit{video and text} tutorial group with a strong effect size ((t) $p_{\mathit{Text-Both}}=0.0257$, Cohen's $d =0.919$, $r=0.417$, $\mathit{MD}=152.407$). Between the other groups there was no statistically significant difference ((w) $p_{\mathit{Video-Text}}=0.65$, (w) $p_{\mathit{Video-Both}}=0.0849$).

The results in Fig. \ref{fig:duration_T2} show that for the scenario tutorial, there were significant differences as well. This time the video tutorial took longer than the text tutorial with a very strong effect size ((t) $p_{\mathit{Video-Text}}=0.0005$, Cohen's $d=3.14$, $r=0.843$, $\mathit{MD}=242.571$). Also, the \textit{video and text} tutorial group took longer than the text tutorial group with a very strong effect size ((t) $p_{\mathit{Text-Both}}=3.4722 \cdot 10^{-5}$, Cohen's $d=3.441$, $r=0.865$, $\mathit{MD}=285.286$). In this case, there was no significant difference between the video tutorial group and the \textit{video and text} group ((t) $p_{\mathit{Video-Both}}=0.1058$).

The null hypothesis that all three groups need the same time can be rejected in both cases as there are differences between some of the groups.

\paragraph{$\mathbf{H_{2b}}$ -- Duration of Tutorial and Task}
Concerning the time needed for the tutorial and the following task, the results are different. For the SIF tutorial and the task there are no significant differences between the three groups ((w) $p_{\mathit{Video-Text}}=0.793$, (w) $p_{\mathit{Video-Both}}=0.5112$, (w) $p_{\mathit{Text-Both}}=0.6848$). The same result shows for the scenario tutorial and the following tasks ((w) $p_{\mathit{Video-Text}}=0.3176$, (w) $p_{\mathit{Video-Both}}=0.2593$, (t) $p_{\mathit{Text-Both}}=0.724$). Therefore, the null hypothesis that there are no differences between the three groups cannot be rejected.

\paragraph{$\mathbf{H_{3a}}$ -- Looking Things Up}
\label{H3a}
A test confirmed that the values in Fig. \ref{fig:lookingUp} are significantly different with a very strong effect size ((w) $p=0.0067$, Cohen's $d=1.274$, $r=0.547$, $\mathit{MD}=1.907$). The participants looked up more things in the text tutorial than in the video tutorial. The same applies for Fig.~\ref{fig:lookedUpBoth} with a very strong effect size ((w) $p=0.0029$, Cohen's $d=1.7$, $r=0.648$, $\mathit{MD}=1.15$). We can reject the null hypothesis that the participants look things up in both tutorials equally often. They looked up more items in the text tutorial.

\paragraph{$\mathbf{H_{3b}}$ -- Understanding}
We thought that there is a difference in how much the participants understood when reading/watching the tutorial for the first time. With a p-value of 0.9462 in a Wilcoxon test, we cannot confirm this result. We cannot reject the null hypothesis. There is no significant difference between the two tutorial types.

\paragraph{$\mathbf{H_{3c}}$ -- Error Quota}
Regarding the number of correct and wrong answers for all four tasks (the final task was divided into two parts) we could not find any significant differences between the three groups. Table \ref{tab:test} shows the p-values of the Wilcoxon tests. The null hypothesis cannot be rejected.

\begin{table}[!htp]
	\centering 
	
	\caption{P-values of the significance tests for comparing the results of the playgrounds and the final task}
	\label{tab:test}
	
	\normalsize

\begin{tabular}{ccccc}
	 
	& Playground 1 & Playground 2 & Final task 1 & Final task 2 \\ 
	\hline 
	Video-Text &  0.3531 & 1 & 0.5774 & 1 \\ 
	 
	Video-Both & 0.3531 & 0.3914 & 0.5774 & 0.5774 \\ 
	 
	Text-Both & 1 & 0.3914 & 1 & 0.5774 \\ 
	 
\end{tabular} 

\end{table}

\paragraph{$\mathbf{H_{3d}}$ -- Level}

We checked with a t-test if there are significant differences in terms of the level between the three tutorial types or between the groups with the scenario task and the groups without it. Against our expectations, there are no significant differences in a Wilcoxon test ($p_{A-B}=1$, $p_{A-C}=0.7195$, $p_{B-C}=0.6589$, $p_{D-E}=0.3431$, $p_{D-F}=0.4744$, $p_{E-F}=0.6812$, $p_{A-D}= 0.1023$, $p_{B-E}=0.7195$, $p_{C-F}=0.467$). We cannot reject the null hypothesis.

\section{Discussion} 
\subsection{Evaluation of Results and Implications}
\paragraph{$\mathbf{H_1}$ -- Usage of the Tutorials}
Contrary to our hypothesis, there was no difference to be found regarding how often the different tutorial types were chosen. Apparently, the participants had no significant preference for one tutorial type. In addition, we could observe during the experiments that many participants started with the video tutorial and then flipped through the text. This is similar to Palmiter and Elkerton~\cite{palmiter} who concluded ``Even with accompanying spoken text, the simplicity of using animated demonstrations may encourage superficial processing and disregard for the procedural text.''

\paragraph{$\mathbf{H_{2a}}$ -- Duration of Tutorial Usage}
For the SIF tutorial, the groups with both tutorials took longer to get through the tutorial. This might be because they had two sources and therefore had a look into both which, of course, takes longer. \\
There was no difference between text and video in this case, which might be because the whole program was new to the participants and, therefore, the participants with the text tutorial took their time to work through the text and fully understand everything. 

For the scenario tutorial, again the group with both tutorials took longer than the text tutorial group, for the same reason. Additionally, for this tutorial the video group also took longer. This can have two reasons: either the text tutorial group was faster because they could easily skip uninteresting parts or because the scenario video took longer than the previous video and now the time difference was bigger.  

All in all, we conjecture that the readers are faster because they can choose how fast they read. The video is always of the same length. The longer the video, the bigger the difference between video and text.

\paragraph{$\mathbf{H_{2b}}$ -- Duration of Tutorial and Task}
We did not expect the results for this hypothesis at all. Although the text tutorial group was faster during the tutorial itself, they were not faster overall, including the following task. We have one explanation for this. As the participants looked up more things in the text tutorial, our conclusion is that the text tutorial group might not have understood as much as the other groups and therefore needed more time during the task to look up information. So the time saved due to fast reading is compensated for by looking up things more often.

\paragraph{$\mathbf{H_{3a}}$ -- Looking Things Up}
Overall, most participants looked up more things in the text tutorial. This might be because it is easier to find information in a text than in a video. Maybe the video was better in explaining and, therefore, the participants did not have to look things up, though this is unlikely, as the participants stated to have understood both tutorial types equally well. Another explanation could be that the participants skipped some parts while reading. Both interpretations are possible.

\paragraph{$\mathbf{H_{3b}}$ -- Understanding}
 We first created the videos and then transcribed them into a text. That is why the given explanations are the same, just the format is different. This leads to the conclusion that it does not matter how an explanation is given, as long as the explanation itself is good. A bad explanation will not get better just because one uses a different format.

\paragraph{$\mathbf{H_{3c}}$ -- Error Quota}
For the practicing tasks as well as for the final task all three groups made about the same number of errors. Contrary to our hypothesis, no tutorial taught the program better, which may have led to a better result in general. As the tutorials were equally good, there is also no difference in the result oft the final task. Hence, our results support the more general results in~\cite{payne, alexander, deVaney}. 

We strongly contradict~\cite{mestre} who found static text as better. Yet, they mostly did a qualitative study on recreating steps from the tutorials. We believe the tasks in our experiment were more complex which might explain why we cannot see a difference here. Furthermore, they found some relationship to the the visual or auditory learning style. Potentially, this could be an influence that we have not controlled. 

In addition, our results are also contrary to related work on the advantages of video and animated tutorials. Baecker~\cite{baecker} makes a case for moving pictures for learning but also states ``Yet the evidence is not conclusive.''  Van der Meij and van der Meij~\cite{vanderMeij} showed significant improvements with video tutorials. One reason might be the difference in age. Their participants had a mean age of 11.8 while ours were all around 20 years old. Furthermore, they also state ``One limitation concerns the absence of measures of transfer. Both the training tasks and the post-test assignments were highly similar to the tasks on which the users were instructed. Because of the ultimate aim of the training is to give the user the capability to complete a range of formatting tasks, future studies should investigate wether the instructions yield any transfer to untrained but related formatting tasks and whether users receiving video instructions also do better on these tasks.'' To some degree we have done that in this experiment. The experiment task was different and complex for the participants. Finally, Lloyd and Robertson~\cite{lloyd} found that students learned statistics better with screencast tutorials. Here, the context might play a role again. They were
not learning a software tool but statistical concepts. It might be that for more abstract concept learning good dynamic visualisations are more helpful than
for software tools.

Therefore, we also conclude for our context of developers learning software tools that the format of a tutorial is less important than the content. As the content was exactly the same, the results were the same as well. The many wrong results in the final task can be explained by the not that good Excel knowledge of many participants. The playgrounds show that it was possible to solve the tasks no matter which tutorial was used but for the more difficult final task the lack of Excel knowledge caused problems for some participants. 

\paragraph{$\mathbf{H_{3d}}$ -- Level}

Quite unexpected, having to go through the advanced tutorials with additional information to learn had no impact on how easy or difficult the participants perceived the experiment to be. We see two possible explanations here: either the format of the tutorial really makes no difference or the tasks and the technology under evaluation were too easy or too hard for a difference to become evident.

\subsection{Threats to Validity}

We have identified a number of threats to validity for our study and split them into three groups: construct validity (CV), internal validity (IV) and external validity (EV). We discuss them in the following:

\begin{itemize}
\item (CV) We took the opinion of the participants directly as a measurement. Still, it might be that some participants misjudged their usage of the tutorial. Nevertheless, this should balance out due to the number of participants. 
\item (CV) As the participants had the choice between three different experiments, the results cannot be mapped to computer science students from our university in general since probably students who are interested in Microsoft Excel actively decided to take our experiment.
\item (CV) While the authors of the tutorials are experienced lecturers, they are not professional tutorial creators. Therefore, our tutorials might differ from those created by professional tutorial creators. 

\item (IV) There might be further personal differences in cognition, personality or learning style~\cite{mestre} in the participants that would explain 
preferences and potentially the duration and error quota. We have not controlled such factors.

\item (IV) The participants had access to the Internet during the experiment. Potentially, this could give a participant additional information that could
distort the results. Only a single participant tried to find information specifically about the Spreadsheet Inspection Framework but did not even find its website.
Other searches on formulas were done by most of the participants and brought them on the same level of Excel understanding. Hence, we believe this
actually improves internal validity. At the same time, by providing a more realistic scenario, we improved external validity. 

\item (IV) We did not have a control group that did not use any tutorial at all. Hence, we cannot show that any of the tutorials improved learning in
comparison to no tutorial. As recommended by Prechelt~\cite{prechelt}, avoiding frustration in experiments is very important. We decided that giving the participants no tutorial would frustrate them because the investigated tool is not self-explanatory.  Taking the decision to experiment without a control group in such settings has also been done in the study by van der Meij and van der Meij~\cite{vanderMeij}. Furthermore, Payne et al.~\cite{payne} already showed an improvement of using animated demonstrations over having no tutorial.

\item (EV) Although we had a total of 42 participants, the sample size is not sufficient for generalizing our findings to all contexts in which tutorials are used. Yet, we found several statistically significant differences with medium to strong effect sizes. Hence, we believe that in cases where software tools shall be learned by developers and there is a choice between a text and a (non-interactive) video tutorial, the time needed for understanding the text tutorial will be significantly shorter. Yet, from our results, we would not expect a significant difference in the overall performance in later tasks or errors made.
\item (EV) Instead of using experienced software developers we used software engineering students in our study. However, if even rather unexperienced students get along with a tutorial, we assume that the tutorial should be suitable for experienced professional developers as well.
\item (EV) While for the purpose of our experiment it was productive to have a text tutorial which is equivalent to a video tutorial content-wise, in practice a combination with one tutorial type covering basic concepts supplemented by another tutorial type covering advanced topics might work even better. 
\item (EV) The study only investigated short-term learning effects, since the participants had to apply the learned content directly after consuming the tutorial. However, there could be significant differences when comparing long-term learning effects.
\item (EV) The spreadsheets we provided to the participants contained seeded errors which is problematic as Panko explains~\cite{panko2}. While this might be regarded as a general threat for studies, we do not see a negative impact in the context of this concrete study.
\item (EV) We only used content-equivalent representatives for two text and video tutorials which explain a single software tool and which were produced by the same authors. To further generalize our findings, it would be necessary to investigate more content-equivalent tutorials.
\end{itemize}

Overall, the answers we found seem reasonable to us. We confirmed them by the aforementioned statistical tests (t-test, Shapiro-Wilk test and the Wilcoxon signed-rank test) and found no contradictions. Nevertheless, we encourage other researchers to replicate our experiment to further confirm our results.  SIF is available at the URL provided in~\cite{sifei}. The raw data we used can be found at~\cite{kulesz_daniel_2016_188896}.

\subsection{Lessons Learned}

As described in section \ref{sibsec:deviations} we lost one set of data. We had taken precaution to not use an account twice by changing the account's icon as a warning that the account was already used. Unfortunately, two experiments took place right after each other so one of us forgot to change the icon and so the account was used twice. For the future, we recommend to save every data set right after the experiment and to schedule a longer break between the time slots. Additionally, we would rephrase the question \enquote{Which tutorial did you use more often} in Fig. \ref{fig:moreOften} in a way that a Likert scale is more applicable.

Regarding the experiment, we were surprised that so many computer science students had so little knowledge of Excel. This did not influence the perception of the tutorials but clearly influenced the many wrong results in the final task. For further experiments we would choose either selected participants with a broader knowledge of Excel or tasks which require less previous knowledge of Excel. We would also recommend to give a cheat sheet about the more unknown formulas used in Excel. Many participants did not know the VLOOKUP formula and therefore had to look it up which took time. With a cheat sheet and more skilled participants there would be no need for internet access.

\section{Conclusions and Future Work}

\subsection{Summary} 
In this study we wanted to investigate differences in educational effects between video tutorials and text-transformed-from-video tutorials for developers learning software tools. 
In our sample, there was no clear preference for static text or non-interactive video tutorials. What we could see is that when provided with both, many participants first watched the video to get a general overview and then used the text to look things up. 
Regarding the time needed to complete the tutorials, we could see that the groups with both tutorials needed more time than the other two groups and also that for the longer scenario tutorial the video group was slower than the text group. 
There was no difference in how much the participants understood after reading/watching the tutorials for the first time, but if they had to look something up they used the text tutorial. We could not see any significant differences in the amount of right and wrong answers to the tasks which leads to the conclusion that both tutorial types taught the features of SIF equally well to our participants. There was also no difference in how easy or hard the participants thought the study was; therefore, we think that the amount of taught features does not influence the perception of the tutorials.

This leads to the conclusion that in an educational context, if tutorials are provided at all, it would be the best for software makers to provide developers with both text and video tutorials. As we assume that authoring text tutorials takes less effort than authoring comparable video tutorials, one could argue that text tutorials provide much better value. However, most developers prefer watching videos in the beginning instead of reading text. Therefore, software makers who only provide text tutorials will need to spend more effort to motivate developers to actually consume their text tutorial.

\subsection{Future Work} 
There are many more different tutorial approaches than just text and video tutorials, e.g. interactive tutorials (though most of them are for games and fewer for application software). For further research it would be interesting to see how these tutorial types compare to our tutorials.  Just because videos and text tutorials seem to be most prominent, this does not mean the lesser known and therefore lesser used methods have to be less effective. 

Also, replicating our study with more participants or with typical end-users without a background or previous knowledge in software engineering could help to generalize our results to other contexts. Therefore, it would be interesting to see how users from other fields work with unknown software, what problems they have and if they generally have a different approach towards tutorials than developers.

Another aspect might be the actual activity at which a tutorial type is targeted. Is it targeted just at teaching the basics or rather advanced topics?

As there were no significant differences regarding the results we think that it might indeed be a personal choice what someone prefers. It would be interesting to see if other characteristics of a person influence this personal preference, e.g. the learning style or the knowledge of the language that is used. It might be easier to understand a video where one can see what is clicked and done than a written text if one is not that familiar with the used language. 

Finally, merging the results of previous studies (indicating general helpfulness of tutorials) with our study's findings, we think that economic aspects should be taken into account as well: there is little to no insight about the effort it takes \enquote{professionals} to produce these tutorials. We are certain that such economic aspects play a major role for decision-makers.

%% file: content/appendix.tex
\section{Experimental Material}

\subsection{Used Tutorials}
\label{app:tut}

\begin{figure}[h]
	\includegraphics[width=\textwidth]{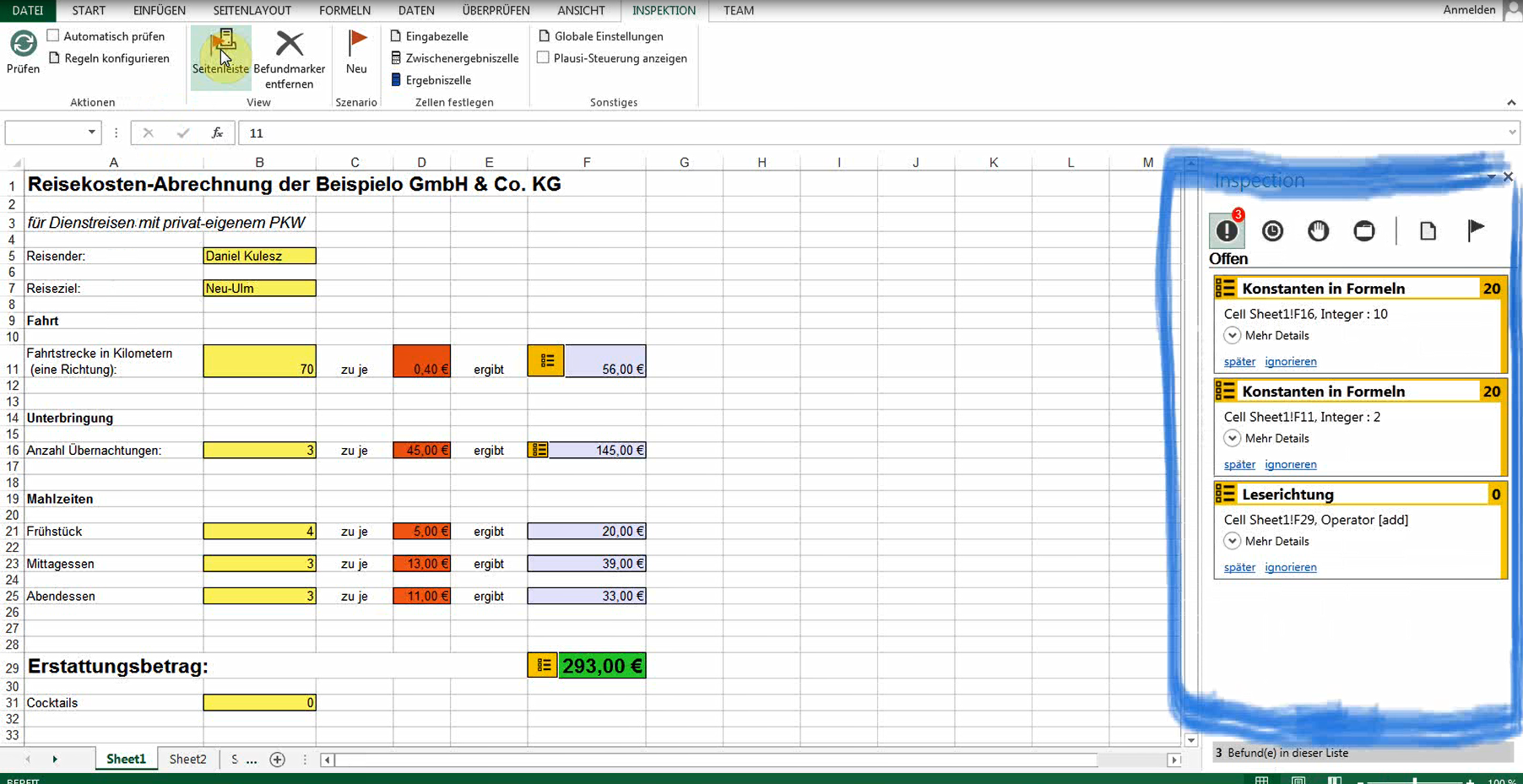}
	\caption{Screenshot of the video tutorial for SIF}
	\label{fig:video}
\end{figure}

\begin{figure}
	\fbox{\includegraphics[clip, trim=2cm 10cm 1.8cm 1.8cm, width=\textwidth]{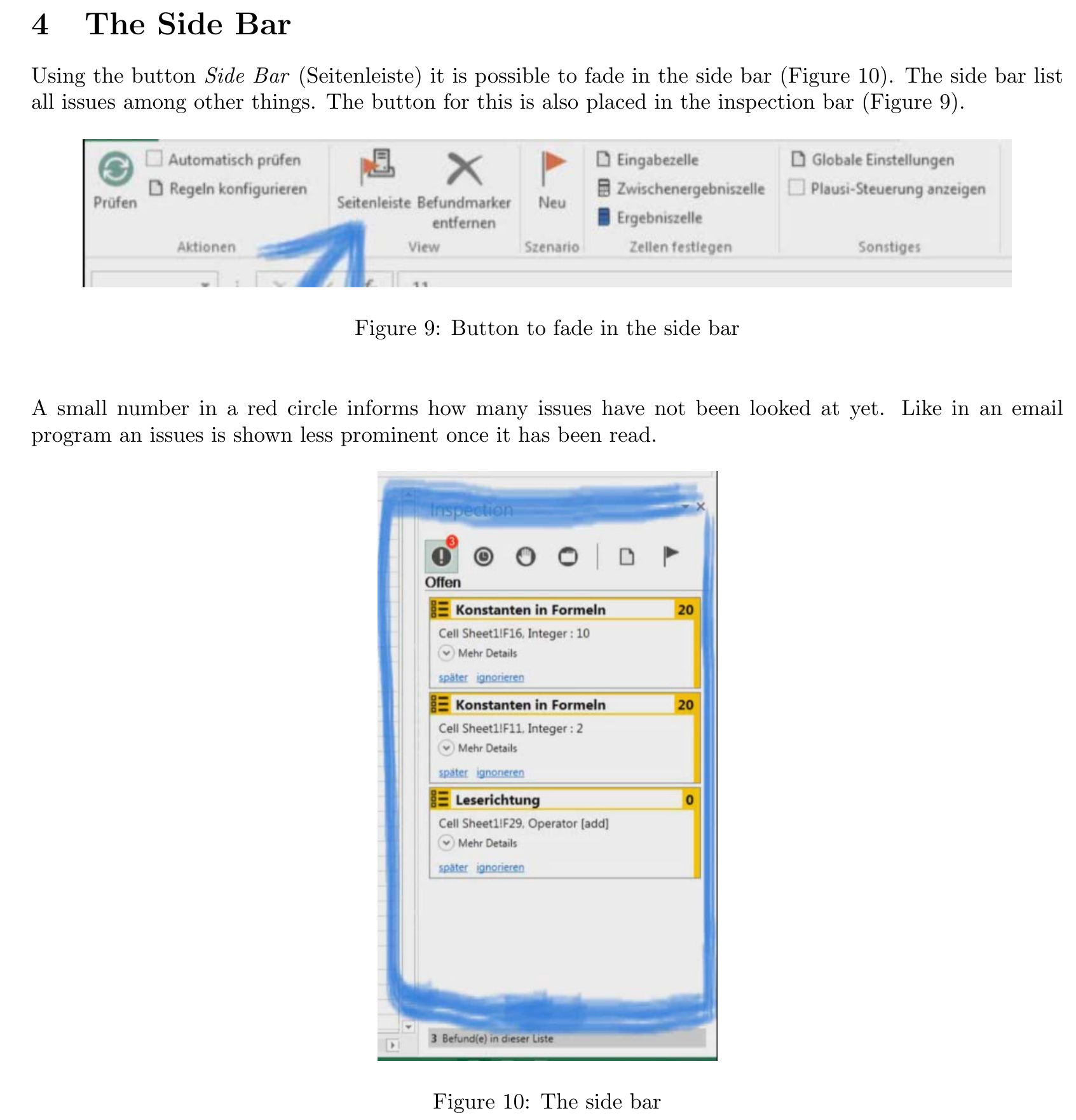}}
	\caption{Excerpt of the text tutorial for SIF (translated)}
	\label{fig:leiste}
\end{figure}

\newpage
\subsection{Pre-Test}
\label{app:pretest}

\begin{center}
	\fbox{\includegraphics[clip, trim=0.8cm 0.8cm 0.8cm 0.8cm, width=0.9\textwidth, page=1]{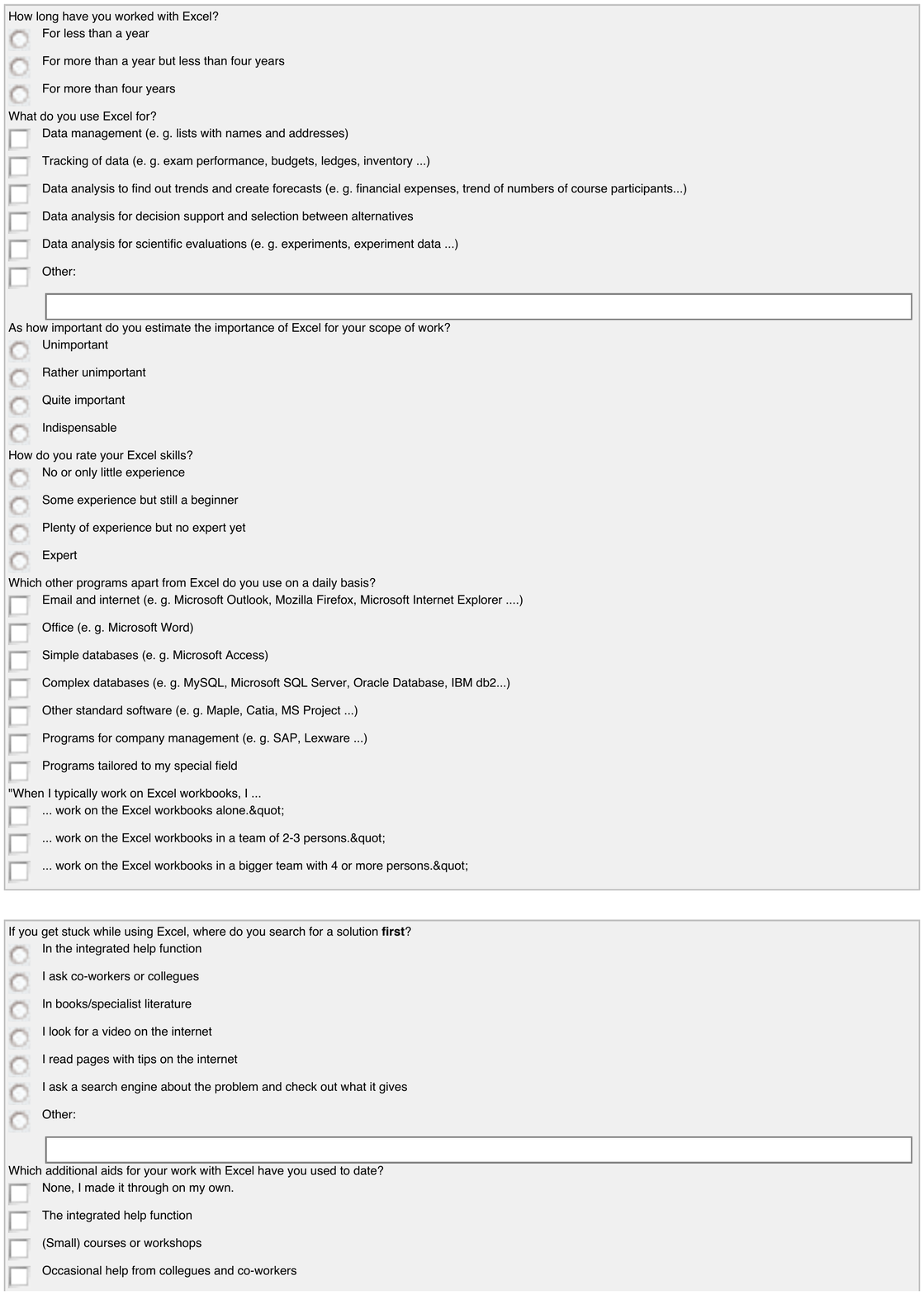}}
	\captionof{figure}{Pre-test, page 1 (translated)}
	\label{fig:pretest1}
\end{center}

\begin{center}
	\fbox{\includegraphics[clip, trim=0.8cm 1cm 0.8cm 0.8cm, width=1\textwidth, page=2]{content/pics/Vorumfrage.pdf}}
	\captionof{figure}{Pre-test, page 2 (translated)}
	\label{fig:pretest2}
\end{center}

\newpage
\subsection{Final Task}

\begin{figure}[h]
  \fbox{\includegraphics[clip, trim=2cm 2cm 1.8cm 12cm, width=\textwidth]{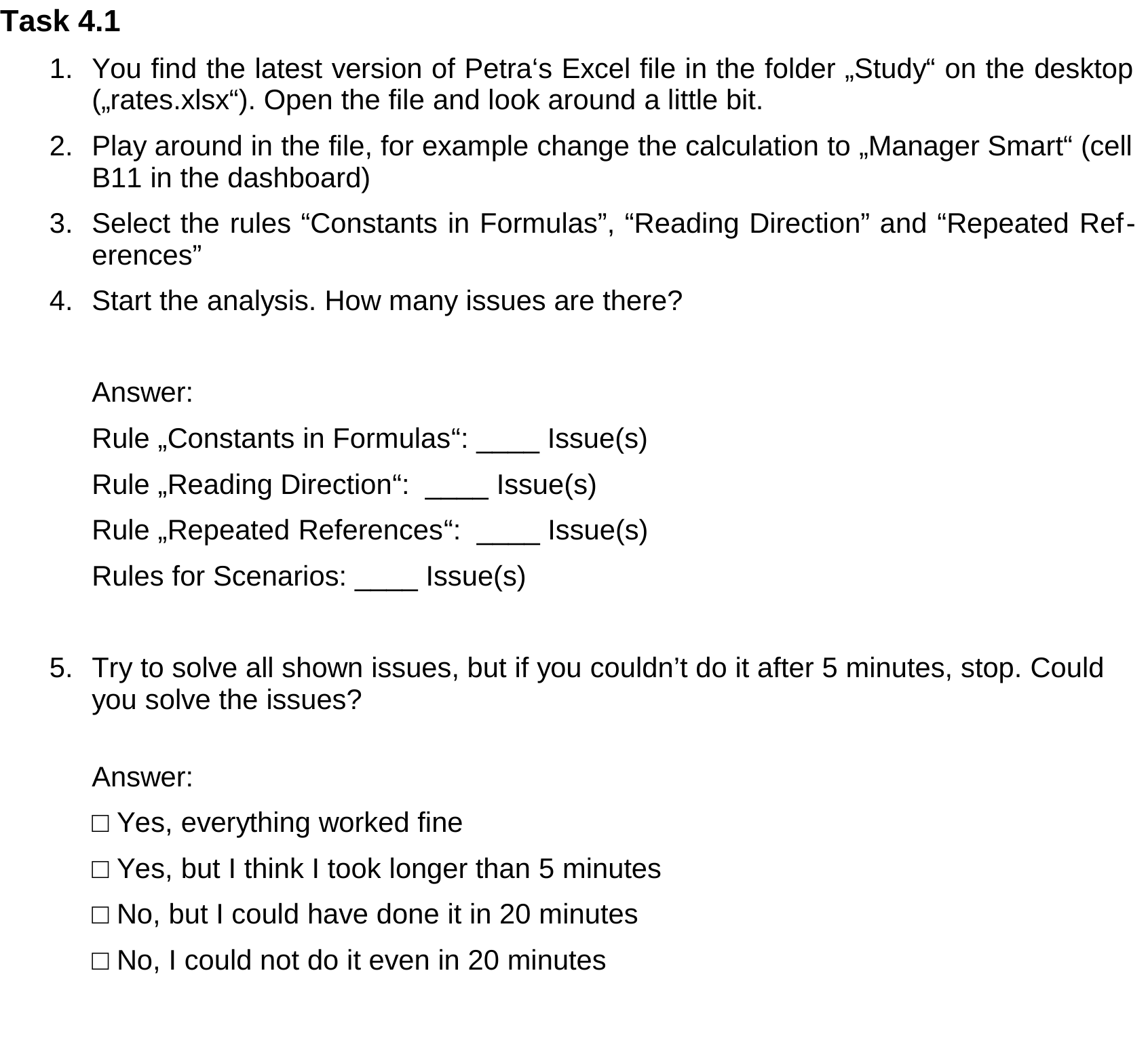}}
  \caption{Translated instructions for the final task}
  \label{fig:instr}
\end{figure}

\newpage

\subsection{Final Test}
\label{app:final_questions}

\begin{center}
  \fbox{\includegraphics[clip, trim=2cm 3cm 1.8cm 1.5cm, width=0.98\textwidth, page=1]{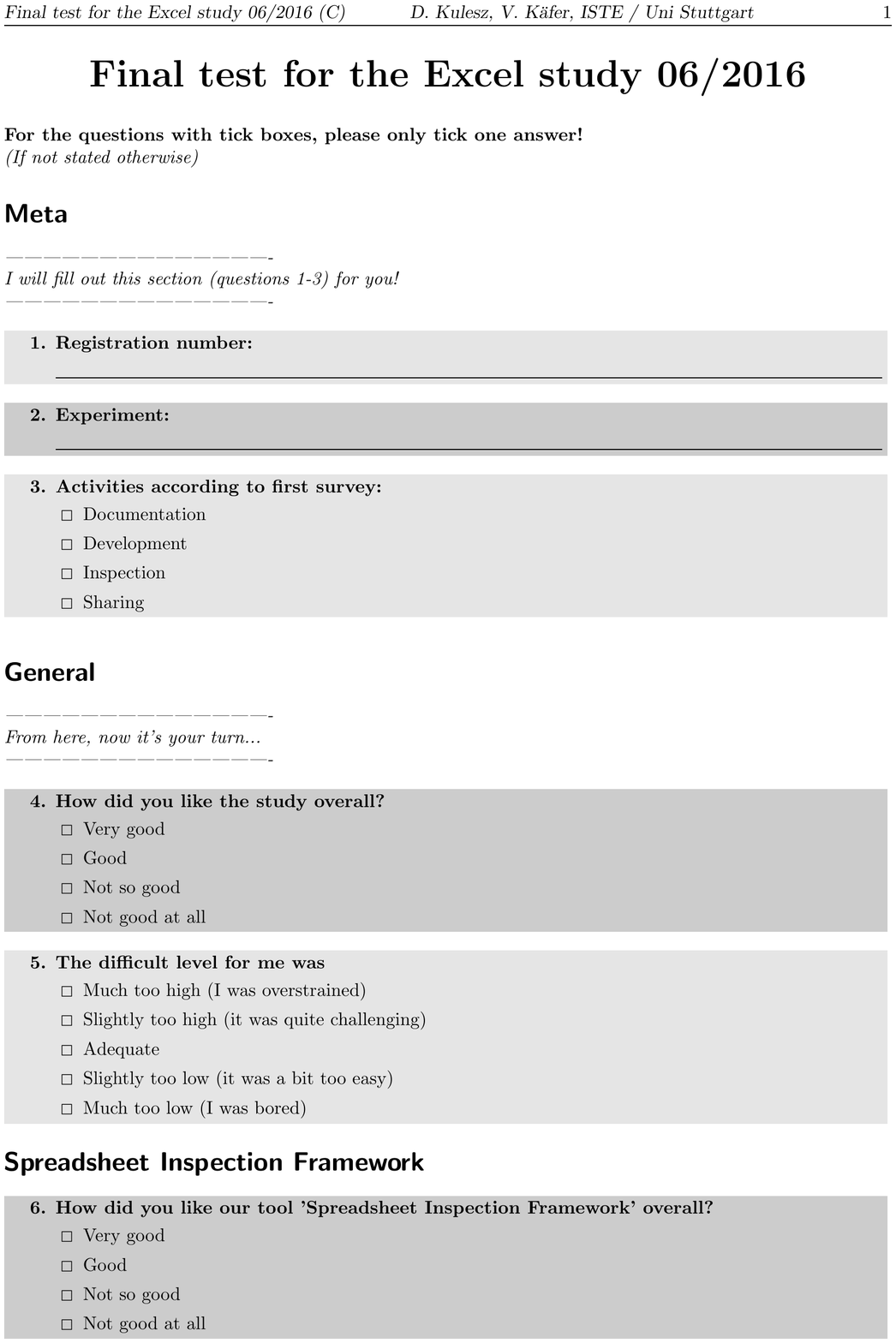}}
  \captionof{figure}{Final test, page 1 (translated)}
  \label{fig:final_questions1}
\end{center}

\newcounter{ct}
\forloop{ct}{2}{\value{ct} < 9}%
{%
\begin{center}
    \fbox{\includegraphics[clip, trim=2cm 1.8cm 1.8cm 1.5cm, width=0.95\textwidth, page=\value{ct}]{content/pics/abschluss-C.pdf}}
    \captionof{figure}{Final test, page \thect ~(translated)}
    \label{fig:final_questions_rest}
\end{center}
}

\section{Miscellaneous}
\begin{figure}[h]
  \includegraphics[width=\textwidth]{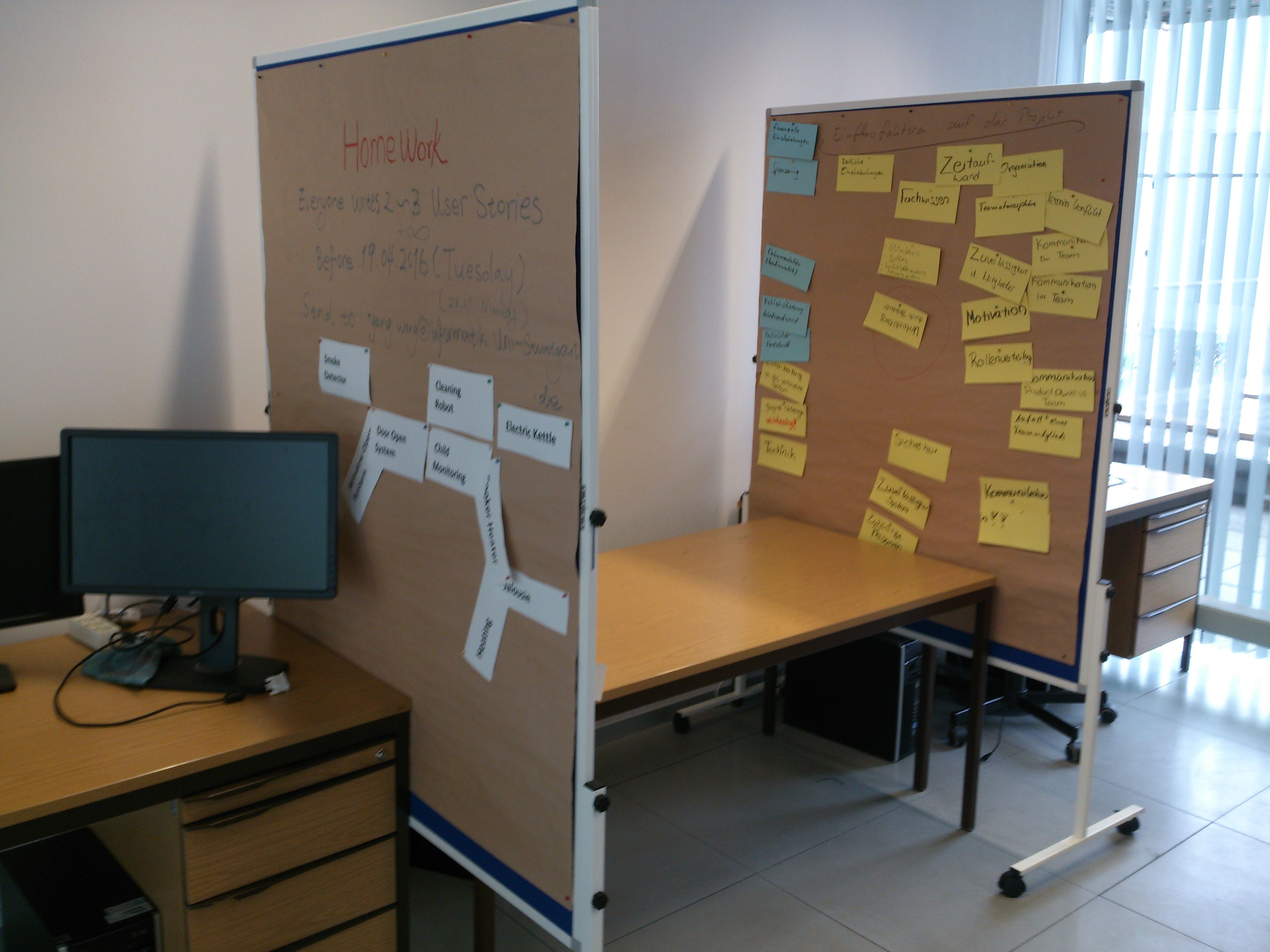}
  \caption{Our computer pool where the experiments took place}
  \label{fig:pool}
\end{figure}